\documentclass[twocolumn,english,aps,floats,pra,noshowpacs,superscriptaddress]{revtex4-1}
\usepackage[T1]{fontenc}
\usepackage{units}
\usepackage{amsmath}
\usepackage{amssymb}
\usepackage{graphicx}
\usepackage{esint}

\makeatletter
 
 \@ifundefined{textcolor}{}
 {%
   \definecolor{BLACK}{gray}{0}
   \definecolor{WHITE}{gray}{1}
   \definecolor{RED}{rgb}{1,0,0}
   \definecolor{GREEN}{rgb}{0,1,0}
   \definecolor{BLUE}{rgb}{0,0,1}
   \definecolor{CYAN}{cmyk}{1,0,0,0}
   \definecolor{MAGENTA}{cmyk}{0,1,0,0}
   \definecolor{YELLOW}{cmyk}{0,0,1,0}
 }

\usepackage{epsfig}
\usepackage{times}
\usepackage{graphics}
\usepackage{dcolumn}
\usepackage{bm}
\usepackage{epic}
\usepackage{eepic}
\usepackage{float}
\usepackage{multirow}
\usepackage{rotate}
\@ifundefined{definecolor}{\usepackage{color}}{}

\makeatother

\usepackage{babel}
\begin{document}

\title{Nonequilibrium phases in hybrid arrays with flux qubits and NV centers}

\author{Thomas H\"{u}mmer}

\affiliation{Instituto de Ciencia de Materiales de Arag\'{o}n y Departamento de
F\'{\i}sica de la Materia Condensada, CSIC-Universidad de Zaragoza,
E-50009 Zaragoza, Spain.}

\affiliation{Institut f\"{u}r Physik, Universit\"{a}t Augsburg, Universit\"{a}tsstra\ss{}e~1,
D-86135 Augsburg, Germany}

\author{Georg M. Reuther}

\affiliation{Institut f\"{u}r Physik, Universit\"{a}t Augsburg, Universit\"{a}tsstra\ss{}e~1,
D-86135 Augsburg, Germany}

\author{Peter H\"{a}nggi}

\affiliation{Institut f\"{u}r Physik, Universit\"{a}t Augsburg, Universit\"{a}tsstra\ss{}e~1,
D-86135 Augsburg, Germany}

\author{David Zueco}

\affiliation{Instituto de Ciencia de Materiales de Arag\'{o}n y Departamento de
F\'{\i}sica de la Materia Condensada, CSIC-Universidad de Zaragoza,
E-50009 Zaragoza, Spain.}

\affiliation{Institut f\"{u}r Physik, Universit\"{a}t Augsburg, Universit\"{a}tsstra\ss{}e~1,
D-86135 Augsburg, Germany}

\affiliation{Fundaci\'{o}n ARAID, Paseo Mar\'{\i}a Agust\'{\i}n 36, 50004 Zaragoza,
Spain}
\begin{abstract}
We propose a startling hybrid quantum architecture for simulating
a localization-delocalization transition. The concept is based on
an array of superconducting flux qubits which are coupled to a diamond
crystal containing nitrogen-vacancy~(NV) centers. The underlying
description is a Jaynes-Cummings-lattice in the strong-coupling regime.
However, in contrast to well-studied coupled cavity arrays the interaction
between lattice sites is mediated here by the qubit rather than by
the oscillator degrees of freedom. Nevertheless, we point out that
a transition between a localized and a delocalized phase occurs in
this system as well. We demonstrate the possibility of monitoring
this transition in a non-equilibrium scenario, including decoherence
effects. The proposed scheme allows the monitoring of localization-delocalization
transitions in Jaynes-Cummings-lattices by use of currently available
experimental technology. Contrary to cavity-coupled lattices, our
proposed recourse to stylized qubit networks facilitates (i) to investigate
localization-delocalization transitions in arbitrary dimensions and
(ii) to tune the inter-site coupling \textit{in-situ}.
\end{abstract}

\date{\today}

\maketitle
\global\long\def\hc{\operatorname{h.c.}}

\section{Introduction}

In recent years, a variety of novel experimental approaches have enabled
tests of fundamental quantum physics such as superpositions, entanglement,
tunneling or quantum phase transitions in artificial devices. Prominent
examples hereof are quantum circuits, quantum dots or optical lattices~\cite{you2011atomicphysics,buluta_natural_2011}.
Apart from their fundamental relevance, quantum technologies allow
realizing quantum information processors that bring along the potential
of carrying out specific tasks at exponentially reduced computation
time~\cite{ladd2010quantum}. Furthermore, quantum simulators of
Feynman type~\cite{feynman1982simulating,houck2012onchipquantum},
employed to simulate the dynamics of one quantum system by means of
another one are a pivotal example of quantum speed-up as compared
to a classical computer. While both systems share the same dynamics,
the simulator offers far more configurability and is better accessible
for a measurement.

Generally, the possibility of observing quantum effects strongly depends
on the coherence properties of the underlying system. With regard
to superconducting circuit qubits, the advantages of versatile manufacturing,
detection and manipulation are paid for at the price of quite high
decoherence rates as compared to trapped ions or spin qubits. One
way out is given by the recently emerging field of hybrid systems~\cite{amsuss2011cavityqed,kubo2011hybridquantum,kubo2010strongcoupling,schuster2010highcooperativity,imamoglu2009cavityqed,rabl2006hybridquantum,verdu2009strongmagnetic,wesenberg2009quantum,xiang2012hybridquantum}.
The main motivation for building hybrid systems is to combine two
advantages: the addressability of artificial quantum circuits and
the long coherence times of elemental systems such as nitrogen-vacancy~(NV)
centers in diamond or in polar molecules. Usually these hybrid systems
are motivated by using the natural spins for building quantum memories.

In this work, we point out an alternative application for the exploration
of many-body physics such as quantum phase transitions. In particular,
we investigate the localized and delocalized phases that occur in
Hubbard-like models such as the Jaynes-Cummings (JC) lattice~\cite{hartmann2006strongly,greentree2006quantum,angelakis2007photonblockadeinduced,hartmann2008quantum}.
In the localized phase, excitations are localized at individual lattice
sites, whereas they are delocalized across the lattice in the delocalized
phase. We propose an intriguingly simple layout for simulating a JC-lattice:
We use the combination of an already experimentally well-proven flux
qubit array together with a single large NV-center crystal. By means
of numerical studies, we corroborate that this system exhibits localized
and delocalized phases. Finally, we demonstrate that these phases
can be identified by monitoring the signatures of the non-equilibrium
system dynamics in presence of decoherence and dissipation upon employing
experimentally accessible parameters.

\begin{figure}[!t]
\includegraphics[width=1\columnwidth]{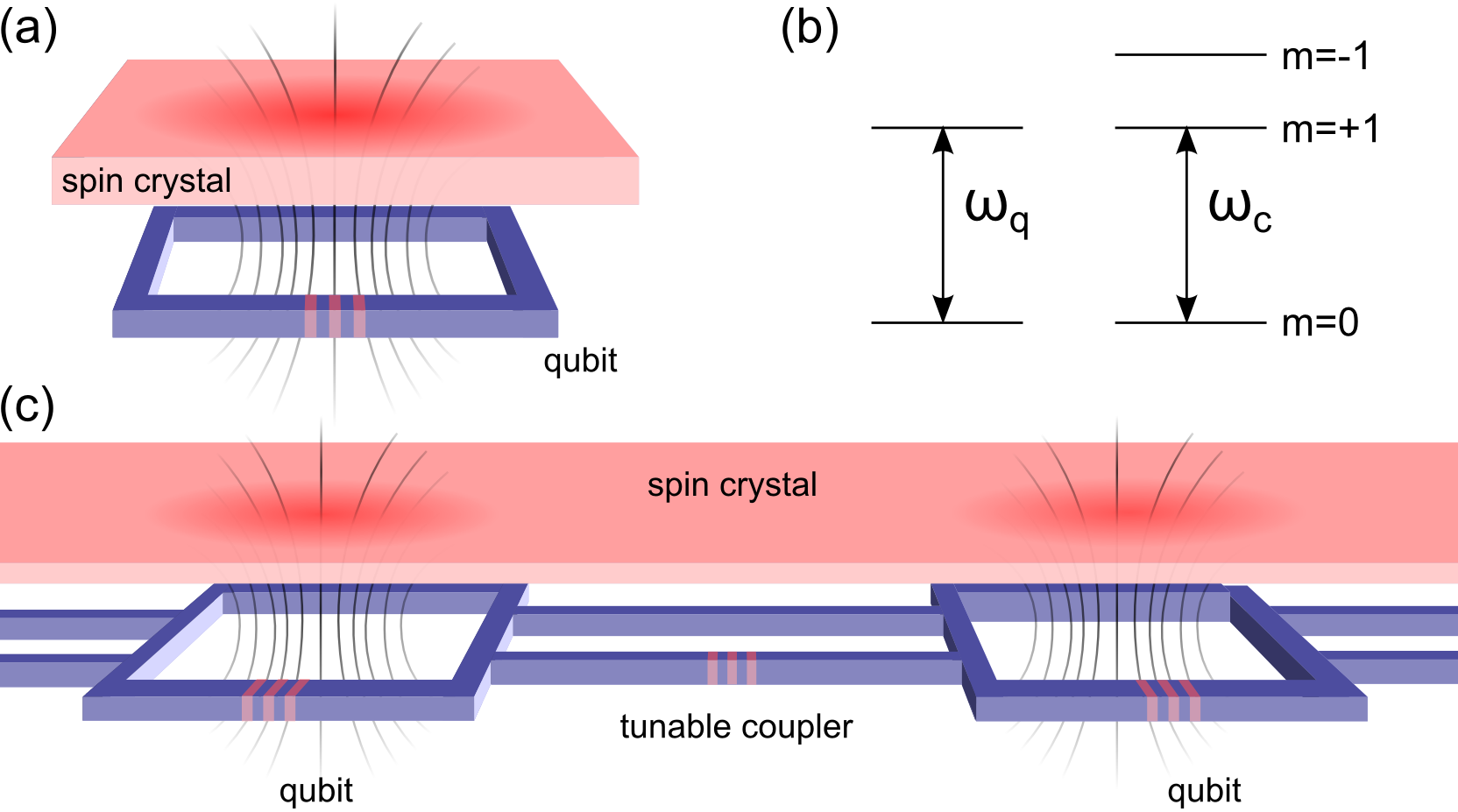} \caption{(a) Schematic of a single flux qubit coupled to a diamond crystal
with NV-centers embedded. (b) Energy diagram for the qubit with level
splitting $\omega_{q}$ and a NV-center. By applying an external magnetic
field, the level with spin projection $m=\pm1$ becomes resonant with
the upper qubit level. (c) The JC-array with tunable qubit-qubit coupling.
Here, adjacent qubits are connected via auxiliary tunable coupler
qubits. Each of the qubits couples to spatially separated regions
of the crystal. The coupler qubit does not couple to the spins because
it is far detuned from the qubits and consquently from the NV-spins
as well.\label{setup}}
\end{figure}

\section{Hybrid qubit-resonator model }

As the elementary unit for the JC-array we propose a hybrid combination
of a flux qubit and an ensemble of independent spins, given here by
the NV-centers in a diamond~\cite{marcos2010coupling,twamley2010superconducting}.
As illustrated in Fig.~\ref{setup}(a), the spin crystal is placed
in proximity of a qubit loop. The spin-qubit Zeeman interaction is
mediated by the magnetic field that stems from the qubit's persistent
currents. A weak external field splits the spin degeneracy so as to
shift one spin transition into resonance with the qubit's transition
frequency $\omega_{q}$. Reducing the spin to its two lowest levels
with mutual energy spacing $\omega_{c}$ (see Fig.~\ref{setup}~(b))
and applying the rotating wave approximation (RWA) we can express
the Hamiltonian of an ensemble of $N$ spins coupled to a single flux
qubit at its degeneracy point as~($\hbar=1$)
\begin{equation}
H_{q+s}=\omega_{q}\sigma^{+}\sigma^{-}+\omega_{c}\,\sum_{k}^{N}\tau_{k}^{+}\tau_{k}^{-}+\sum_{k}^{N}\left(g_{k}\tau_{k}^{+}\sigma^{-}+\hc\right)\,.\label{eq:H}
\end{equation}
Here, $\sigma^{\pm}=\sigma^{x}\pm i\sigma^{y}$ and $\tau_{k}^{\pm}=\tau_{k}^{x}\pm i\tau_{k}^{y}$ are raising and lowering
operators with respect to the qubit's ($\sigma$) and the spins' ($\tau_{k}$) Pauli matrices. The coupling strengths
$g_{k}$ between the qubit and the individual spins are proportional
to the magnitude of the qubit's field at the spin positions~\cite{marcos2010coupling,twamley2010superconducting}.
Assuming the case of a large spin ensemble with low polarization,
i.e., close to its ground state, we introduce a collective operator:
i.e. with $g=(\sum_{k}^{N}|g_{k}|^{2})^{1/2}$ we set $a^{\dagger}=g^{-1}\sum_{k}^{N}g_{k}\tau_{k}^{+}$
together with its hermitian conjugate $a$, yielding approximately
the bosonic commutation relation, $[a,a^{\dagger}]\cong1$ (See Appendix
A). Thus, we interpret the ensemble as an effective \emph{bosonic}
mode and arrive at the effective Jaynes-Cummings model, reading as
\begin{equation}
H_{\textrm{JC}}=\omega_{q}\sigma^{+}\sigma^{-}+\omega_{c}a^{\dagger}a+g\left(a^{\dagger}\sigma^{-}+a\sigma^{+}\right)\,.\label{eq:HJC}
\end{equation}
It describes a collective harmonic oscillator mode being coupled to
a two-level system with the interaction strength $g$. The collective
coupling $g$ is enhanced by a factor of $\sqrt{N}$ compared to the
root mean square of the individual couplings $g_{k}$, see Appendix
\ref{app:A}. Recent experiments achieved coupling strengths as strong
as $g\approx2\pi\times\unit[35]{MHz}$~\cite{zhu2011coherent}.

\section{The Jaynes-Cummings lattice with qubit-qubit coupling }

A general advantage of superconducting circuits is their scalability
and the rich variety of coupling mechanisms that can be implemented
on a chip. In particular, arrays of flux qubits with tunable coupling
strength between individual qubits have been realized using a SQUID
or ancilla flux qubit~\cite{harrabi2009engineered,hime2006solidstate,plourde2004entangling,niskanen2007quantum,vanderploeg2007controllable}.
In recent experiments manipulating coupling strengths \textit{in situ}
and the engineering of various types of circuit connectivity has become
feasible~\cite{harris2009compound,tsomokos2010usingsuperconducting,harris2010experimental,johnson2011quantum,grajcar2005directjosephson}.
In this work, we restrict ourselves to a chain of qubits with tunable
nearest neighbor interaction. This array of qubits can be readily
turned into a JC-lattice of coupled qubit-oscillator systems by putting
one NV-center crystal on top, as sketched with Fig.~\ref{setup}(c).
As argued above, the spin crystal adds an effective harmonic oscillator
degree of freedom to each site of the array. Apart from the possibility
to tune the coupling between the sites of the lattice, the most appealing
aspect of this hybrid architecture is simplicity. Furthermore, the
harmonic oscillators in the form of the spin crystal exhibit excellent
coherence properties, homogeneous transition frequencies and coupling
strengths, all implemented here within a reduced geometric dimension
as compared to coplanar waveguide resonators.

For well-separated qubits, we can neglect their mutual inductance
as well as the cross-coupling of one qubit to the spin ensemble of
another site, being even one order of magnitude smaller.{} This JC-lattice
with $M$ sites is thus described by
\begin{equation}
H_{\textrm{JCL}}=\sum_{j}^{M}H_{\textrm{JC,}j}+J\sum_{j}^{M-1}\left(\sigma_{j}^{+}\sigma_{j+1}^{-}+\hc\right)\ ,\label{eq:JC-array}
\end{equation}
with the single-site Hamiltonians $H_{\textrm{JC},j}$ given in Eq.~\eqref{eq:HJC}.
Here, $J$ denotes the uniform qubit-qubit coupling strength and the
operators $\sigma_{j}^{\pm}$ describing the creation and annihilation
of a qubit excitation at the $j$-th site.

A subtle but salient difference between our model and previously studied
JC-lattices is in the interaction mechanism between individual lattice
sites. While we propose an inter-site coupling mediated by the qubits,
previous works have dealt with the complementary approach where the
lattice sites interact via the oscillator degrees of freedom, as in
coupled cavities~\cite{hartmann2006strongly,greentree2006quantum,angelakis2007photonblockadeinduced,hartmann2008quantum,leib2010bosetextendashhubbard,makin2009timeevolution}
or in superconducting resonators~\cite{koch2009superfluidmottinsulator,schmidt2010nonequilibrium}.
In the latter case, the coupling part of the JC-lattice Hamiltonian~\eqref{eq:JC-array}
assumes the form $a_{j}^{\dagger}a_{j+1}+{\rm h.c.}$ 

\section{Equilibrium properties of the JC lattice }

For Bose-Hubbard--like models the occurrence of a quantum phase transition
between localized and delocalized phases has been extensively studied~\cite{sachdev2011quantum}.
Analogous transitions have been investigated with polaritons in JC-lattices~\cite{hartmann2006strongly,greentree2006quantum,angelakis2007photonblockadeinduced,hartmann2008quantum,koch2009superfluidmottinsulator,schmidt2010nonequilibrium,leib2010bosetextendashhubbard}.
Here, the term polariton refers to the eigenstates $|n,\pm\rangle$
of the single-site Hamiltonian $H_{\textrm{JC}}$ {[}Eq.~\eqref{eq:HJC}{]}.
The excitation number $n$, being the eigenvalues of the operator
$\mathcal{N}=a^{\dagger}a+\sigma^{+}\sigma^{-}$, are conserved due
to $[H_{\textrm{JC}},\mathcal{N}]=0$. Similarly, the full JC lattice
Hamiltonian~\eqref{eq:JC-array} conserves the total number of excitations
in the lattice. The ground state $|0\rangle$ has no excitations $n{=}0$,
while the states $|n,\pm\rangle$ ($n{>}0$) are each twofold degenerate
with respect to $\mathcal{N}$. If the qubit and the resonator are
in resonance, $\omega_{c}\,{=}\,\omega_{q}$, the polaritonic states
are symmetric~($+$) and antisymmetric~($-$) superpositions $|n,\pm\rangle=(|n-1\rangle|{\uparrow}\rangle\pm|n\rangle|{\downarrow}\rangle)/\sqrt{2}$
of the oscillator Fock states $|n\rangle$ and the qubit ground ($|{\downarrow}\rangle$)
and excited ($|{\uparrow}\rangle$) states, respectively.

\begin{figure}[!t]
\includegraphics[bb=0bp 0bp 247bp 369bp,width=1\columnwidth]{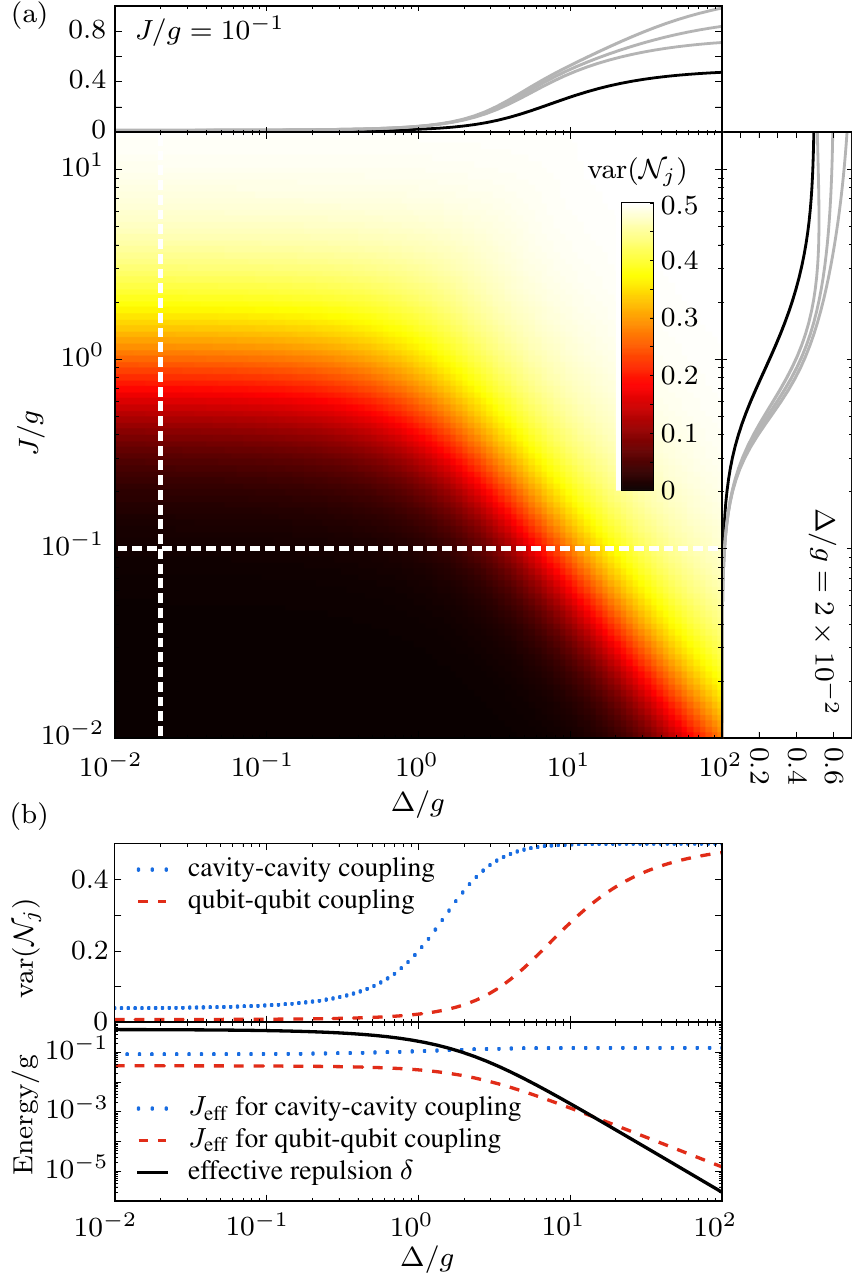}
\caption{Transition between the localized and delocalized phases in a qubit-coupled
JC-array. (a) Fluctuations of the number of excitations at a certain
site ${\rm var}(\mathcal{N}_{j})$ for a two-site setup (both sites
$j=1,2$ yield the same plot) as a function of the inter-site coupling
$J$ and the qubit-ensemble detuning $\Delta$. The dark shaded region
indicates that the system in the localized phase, while the brighter
areas are related to large fluctuations, i.e., the delocalized phase.
The two side panels depict a horizontal cut along $J=0.1g$ and a
vertical cut along $\Delta=2\times10^{-2}g$ (i.e., very close to qubit-oscillator
resonance), respectively. There, the solid black curves depict ${\rm var}(\mathcal{N}_{j})$
for the two-site setup as in the central panel. For comparison, we
have included the fluctuation characteristics ${\rm var}(\mathcal{N}_{j})$
for longer JC-arrays with $N{=}3$--$5$ sites in ascending order,
where $j$ denotes a central site of the array. (b) Comparison of
QQ- and CC-coupled chains. In the latter, the transition occurs at
lower detunings due to higher effective coupling of polaritons between
adjacent sites. This can be seen in the lower plot: Changing the detuning
affects both the effective repulsion $\delta$, as well as the effective
coupling $J_{\textrm{eff}}$. The transition occurs when $\delta$
and $J_{\textrm{eff}}$ cross.\label{fig2eq} }
\end{figure}

The localization-delocalization transition we consider in this work
takes place for the lowest energy state in the subspace with one average
excitation per site~\cite{hartmann2008quantum,angelakis2007photonblockadeinduced}.
As we argue with Appendix \ref{app:conservation}, for weak inter-site
coupling $J$, no inter-conversion between the $+$ and $-$ polaritons
occurs~\cite{koch2009superfluidmottinsulator,angelakis2007photonblockadeinduced}.
Therefore, in order to obtain analytical estimates, we can neglect
$|n,+\rangle$ polaritons and restrict our studies to $|n,-\rangle$
polaritons which are lower in energy. We then introduce the ``effective
repulsion'' $\delta=E_{|2,-\rangle}-2E_{|1,-\rangle}$, i.e.,
\begin{equation}
\delta=-\sqrt{2g^{2}+\frac{\Delta^{2}}{4}}+2\sqrt{g^{2}+\frac{\Delta^{2}}{4}}-\frac{\Delta}{2}\,.\label{delta}
\end{equation}
This positive-valued repulsion increases with the qubit-oscillator
coupling strength $g$ and decreases with the detuning $\Delta=\omega_{q}-\omega_{c}$.
It measures the extra energy needed to insert two polaritons into
a single site as compared to distributing them across two sites. Thus,
a large repulsion promotes an even distribution of excitations over
the lattice sites. In this case, the system eigenstate is approximately given by
a product  of the local single-site eigenstates
$|1,-\rangle_{j}$.

By contrast, a large inter-site coupling quantified by $J$ favors
delocalized excitations, i.e., momentum eigenstates that are given
by a superposition of product states, each with different $n$. Thus,
modifying $J$ or the repulsion (e.g. by means of $\Delta$) one ends
up in two extreme regimes: the localized or the delocalized phase.

As stated at the beginning of this section, the Hamiltonian~(\ref{eq:JC-array})
conserves the total number of excitations. Therefore, the fluctuation
of the excitation number in a particular lattice site, ${\rm var}(\mathcal{N}_{j})=\langle\mathcal{N}_{j}^{2}\rangle-\langle\mathcal{N}_{j}\rangle^{2}$,
is used as an order parameter  in JC-lattices~\cite{koch2009superfluidmottinsulator}.
For ${\rm var}(\mathcal{N}_{j})=0$, the excitations are trapped,
and the system is in a localized phase. By contrast, large fluctuations
indicate the delocalized phase.

In order to investigate the transition between the localized and the
delocalized phases numerically we calculate ${\rm var}(\mathcal{N}_{j})$
for a setup with two sites. As indicated in the main panel of Fig.~\ref{fig2eq}~(a),
a transition between the two different regimes characterized by zero
and finite ${\rm var}(\mathcal{N}_{j})$ occurs upon a change of $\Delta$
or $J$. This two-site setup already exhibits the same qualitative
features as longer arrays of finite length, as we corroborate in the
side panels of Fig.~\ref{fig2eq}(a). There, we compare the variance
for arrays with two or more sites by means of two cross sections through
the main panel for fixed values of $\Delta$ and $J$, respectively.
Thus, the elementary two-site setup -- readily feasible with present-day
experimental techniques -- already allows for a good qualitative estimate
of the transition properties of a JC-array. Experimental feasibility
will be further discussed in section \ref{sec:exp}.

To gain  analytical insight, we express the coupling between
the individual sites in terms of the relevant polaritonic basis states
$|n,-\rangle_{j}$. In doing so, we can approximate $\sigma_{j}^{+}=\sum_{n}^{\infty}s_{n,--}|n+1,-\rangle_{j}\langle n,-|_{j}$,
where the coefficients $s_{n,--}$ depend on $\Delta$, $g$ and $n$
and their explicit form is detailed in Appendix~\ref{app:hopping}.
Thus, two sites initially in the state $|1,-\rangle$ are coupled
with the effective strength (see Appendix \ref{app:hopping})
\begin{equation}
J_{\textrm{eff}}=Js_{0,--}s_{1,--}\;.
\end{equation}
In the lower panel of Fig.~\ref{fig2eq}(b), we compare $J_{\textrm{eff}}$
to the effective repulsion strength $\delta$ {[}Eq.~\eqref{delta}{]},
both plotted as functions of $\Delta$ at fixed $J$. We find that
the observed crossing point of $J_{\textrm{eff}}$ and $\delta$ closely
matches the location of the localization-delocalization transition.

Furthermore, we compare our results for a JC-array with qubit-qubit
(QQ)-coupling to a similar setup with cavity-cavity (CC)-coupling,
i.e., an array in which the individual sites interact via their oscillator
degrees of freedom $J(a_{j}^{\dagger}a_{j+1}+{\rm h.c})$. For this
latter scenario, we find that the transition to the delocalized phase
already occurs at smaller $\Delta$, see Fig.~\ref{fig2eq}(b), top
panel. As in the (QQ)-coupled case we calculate $J_{\textrm{eff}}$
via the relevant polaritonic basis states and indeed find a larger
effective inter-site interaction that hence explains the observed transition
point. The interested reader can check the explicit coupling coefficients
in Appendix~\ref{app:hopping}. With increasing detuning, the $|n,-\rangle_{j}$
polaritons become more and more bosonic, i.e. only the oscillator
degree of freedom is excited, $|n,-\rangle_{j}\approx|{\downarrow}\rangle_{j}|n\rangle_{j}$.
This allows for a simple explanation of the different trends of $J_{\textrm{eff}}$
in Fig.~\ref{fig2eq}(b) when increasing the detuning. In a qubit-coupled
array, the bosonic excitations must hop via the route oscillator-qubit-qubit-oscillator
to reach the next lattice site, therefore $J_{\textrm{eff}}$ is small.
By contrast, in the cavity-coupled setup the bosonic excitations can
hop directly to the next-site oscillator and $J_{\textrm{eff}}$ is
therefore larger when the excitations are purely bosonic.

\section{Non-equilibrium dynamics }

The characterization of localized and delocalized phases at equilibrium
is helpful in exploring the physics in JC-lattices with qubit-qubit
coupling and contrasting it with JC-lattices that interact with a
cavity-cavity coupling. However, the assumption of staying in the
subspace with one mean excitation per site is not completely realistic
in practice. In particular, non-equilibrium processes such as dissipation
and decoherence are crucial in solid state devices. Next we point
out that the signatures of the localization-delocalization transition
remain preserved even in the presence of dissipation. A corresponding
measurement only requires state preparation and qubit readout.

We model dissipation for both the qubit and spin ensemble by means
of a quantum master equation, which for the JC lattice assumes the
form (at zero temperature)~\cite{scala2007cavitylosses,reuther2010tworesonator}
\begin{equation}
\dot{\varrho}(t)=-i[H_{{\rm JCL}},\varrho]+\sum_{j}\left(\gamma_{c}L_{a_{j}}[\varrho(t)]+\gamma_{q}L_{\sigma_{j}^{-}}[\varrho(t)]\right)\;.\label{qme}
\end{equation}
The Lindblad dissipators $L_{O}$ act on the density operator $\rho$
as $L_{O}[\rho]=O\rho O^{\dagger}-\frac{1}{2}(O^{\dagger}O\rho+\varrho O^{\dagger}O)$.
The operators $O=\{\sigma_{j}^{-},a_{j}\}$ describe the system-bath
coupling of the $j$-th qubit and oscillator, respectively, while
$\gamma_{q}$ and $\gamma_{c}$ are the associated, uniform decoherence
rates.

\begin{figure}[!t]
\includegraphics[width=1\columnwidth]{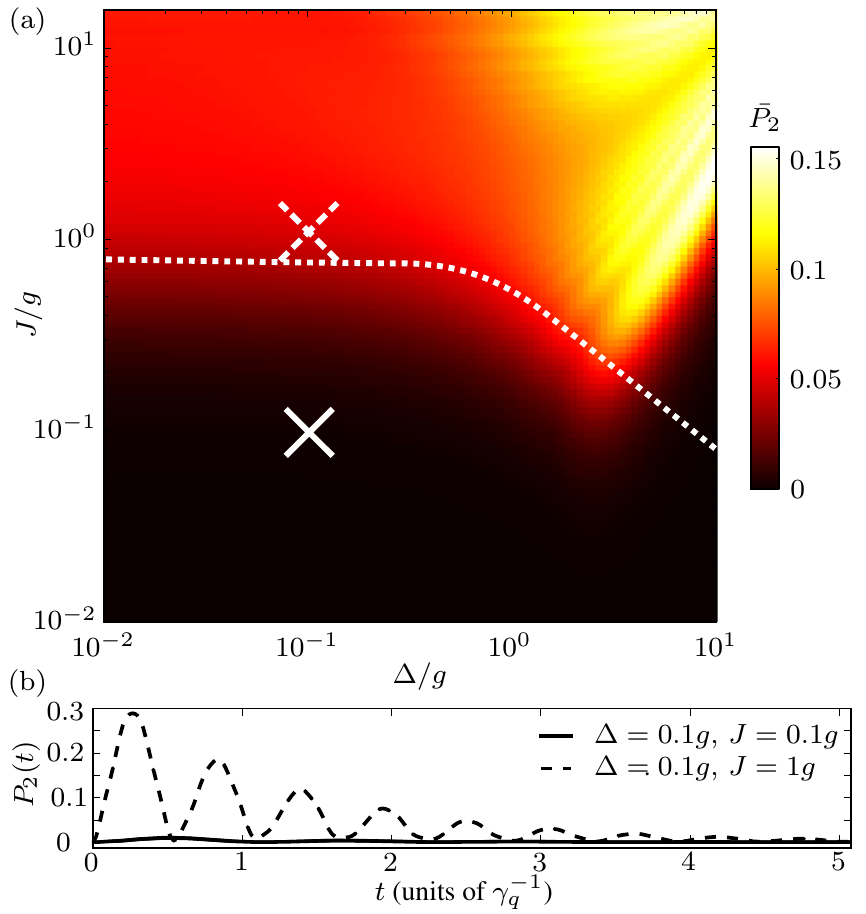}\caption{Non-equilibrium signature of the phase transition. (a) Time-averaged
probability $\bar{P}_{2}$ to find two excitations in a single site.
The coupling is assumed as $g=2\pi\times\unit[10]{MHz}$ and the decay
rates of qubits and oscillators are $\gamma_{\mathrm{q}}=2\pi\times\unit[1]{MHz}$
and $\gamma_{\mathrm{c}}=2\pi\times\unit[0.1]{MHz}$. We choose the
integration time as $T=5\gamma_{\mathrm{q}}^{-1}$. The dotted line
marks the boundary ($\nicefrac{1}{2}\max\{{\rm var}(\mathcal{N}_{j})\}$)
where the phase transition occurs in the equilibrium case of Fig.~\ref{fig2eq}.
(b) Time evolution of $P_{2}$ in two exemplary points in the delocalized
(dashed line and dashed cross in (a)) and localized phase (solid line
and solid cross in (a)), respectively.\label{fig3non-eq} }
\end{figure}

The system is initially prepared with one $|1,-\rangle$ polariton
in each site. Calculating the system dynamics numerically, we obtain
the time-dependent probability $P_{2}(t)=\textrm{Tr}\left\{ \Pi_{2}\varrho(t)\right\} $
of finding two excitations in one site, where $\Pi_{2}=|2\rangle|{\downarrow}\rangle\langle g|\langle2|+|1\rangle|{\uparrow}\rangle\langle e|\langle1|$.
If the system is in the localized phase we expect $P_{2}(t)$ to remain
close to zero. By contrast, in the the delocalized phase, $P_{2}(t)$
reaches finite values over time. This behavior is depicted in Fig.~\ref{fig3non-eq}(b).
Here, it is also visible that the system evolves eventually into its
ground state due to decoherence.

In order to quantify the phase we introduce the averaged probability
\begin{equation}
\bar{P}_{2}=\frac{1}{T}\int_{0}^{T}{\rm d}tP_{2}(t)\,.
\end{equation}
In order to take into account the dynamics before relaxation into
the ground state dominates, the integration time should fulfill $T\gg J_{{\rm eff}}^{-1}$
but $T\lesssim\min\{\gamma_{c}^{-1},\gamma_{q}^{-1}\}$. Fig.~\ref{fig3non-eq}(a)
depicts $\bar{P}_{2}$ as a function of both the hopping parameter
$J$ and the detuning $\Delta$ similar to the equilibrium analysis.
For comparison, the white dashed line marks the parameter regime where
the phase transition occurs in the equilibrium case in Fig.~\ref{fig2eq}.
While we find a good agreement for small values of $\Delta$, the
border between both phases is not resolved in the far-detuned limit
where the effective inter-site coupling strength $J_{\textrm{eff}}$
decreases below the decoherence rates $\{\gamma_{q},\gamma_{c}\}$.

\section{Experimental feasibility }

\label{sec:exp}

We next address the feasibility of the JC array proposed in this paper.
The case of strong coupling between an ensemble of NV-centers ($\omega_{c}\approx\unit[2\pi\times2.88]{MHz})$
and a flux qubit has previously been reported experimentally with
coupling strengths up to $g\simeq2\pi\times\unit[35]{MHz}$ \cite{zhu2011coherent}.
Thus, using our estimations, we  can safely consider a coupling strength of $g\cong2\pi\times\unit[10]{MHz}$.
On the other hand, the experimental accessible tunable qubit-qubit
couplings  are between $2\pi\times\unit[1]{MHz}\leq J\leq2\pi\times\unit[100]{MHz}$~\cite{vanderploeg2007controllable,hime2006solidstate,niskanen2007quantum}.
All together  sets the operation range to $0.1\leq J/g\leq10$.
We next consider the decay rates of the involved subsystems.  Realistic
values for the qubit decay rates are  $\gamma_{q}=2\pi\times\unit[1]{MHz}$
(i.e. $\gamma_{q}/g=0.1$) and for the spin decay rates are $\gamma_{c}\leq2\pi\times\unit[0.1]{MHz}$
(i.e. $\gamma_{c}/g\leq0.01$).

These values were used to obtain the plots in Fig.~\ref{fig3non-eq}.
Therefore, current technology allows for monitoring both phases.
Finally, readout of the number of excitations at a specific site can
be performed by measuring the qubit dynamics.

\section{Conclusions}

In summary, we have introduced a novel JC-lattice based on a hybrid
combination of flux-qubits and NV-centers. In contrast to JC-lattices
based on coupled cavities or superconducting resonators, the harmonic
oscillator degree of freedom (``cavity'') is smaller in size than
the qubit (``atom''). This allows one to couple the individual JC-sites
via the qubits instead of the harmonic oscillators. We have argued
that similarly to \emph{cavity-coupled} JC-lattices a localization-delocalization
transition can be observed in these novel \emph{qubit-coupled} JC-lattices.
Even though localization-delocalization transitions in JC-lattices
have been proposed theoretically some time ago, they could not be
observed in an experiment yet. Our proposal relies on a straightforward
modification of already realized flux qubit arrays by simply mounting
a single NV-center crystal on top. This minimal modification of a
common setup opens the possibility of studying many-body phenomena
in strongly coupled hybrid architectures within state-of-the-art experimental
technology. Apart from its simplicity, further advantages are the
possibility to investigate localization-delocalization transitions
in arbitrary (even fractal) dimensions and to tune the inter-site
coupling in-situ by using common techniques for building flux qubit
networks.
\begin{acknowledgments}
We acknowledge  enlightening discussions with M. Hartmann at TUM. This work
was supported by Spanish MICINN projects FIS2011-25167, CSD2007-046-
Nanolight.es, the European PROMISCE and the DFG via the Collaborative
Research Center SFB 631 and the Nanosystems Initiative Munich (NIM).
\end{acknowledgments}
\appendix

\section{Collective modes}

\label{app:A}

Here we outline the approximation that allows us to express the spin
ensemble by a collective bosonic operator. With homogeneous coupling
for all the spins one can express the spins by a collective angular
momentum operator and then apply a Holstein-Primakoff approximation
to yield a bosonic operator. However, for inhomogeneous couplings
this is not readily possible because the collective operator does
not fulfill angular momentum algebra. Nevertheless, we can arrive
at collective bosonic operators: We start with the Hamiltonian of
spins (inhomogeneously) coupled to a qubit with the individual coupling
strength $g_{k}$
\begin{equation}
H=\omega_{q}\sigma^{+}\sigma^{-}+\omega_{c}\sum_{k}^{N}\tau_{k}^{+}\tau_{k}^{-}+\sum_{k}^{N}\left(g_{k}\sigma^{-}\tau_{k}^{+}+\hc\right)\,.\label{eq:H}
\end{equation}
Here, $\tau_{k}^{\pm}$ are the Pauli raising and lowering operators
of the spins in the ensemble and $\sigma^{\pm}$ the ones of the qubit.
The spins are taken to have the homogeneous energy splitting $\omega_{c}$
and the qubit the splitting $\omega_{q}$. We next introduce the collective
operator
\begin{equation}
a^{\dagger}=\frac{1}{\sqrt{N}\bar{g}}\sum_{k}^{N}g_{k}\tau_{k}^{+}
\end{equation}
where $\bar{g}$ is the root mean square of the individual couplings,
$\bar{g}^{2}\equiv\sum_{k}\left|g_{k}\right|^{2}/N$. In the low polarization
limit, where almost all spins are in the ground state, it follows
that these operators approximately fulfill bosonic commutation relations.
To validate this we calculate the commutator
\begin{align*}
\left[a,a^{\dagger}\right] & =\frac{1}{N\bar{g}^{2}}\sum_{kl}g_{k}^{*}g_{l}\left[\tau_{k}^{-},\tau_{l}^{+}\right]\\
 & =\frac{1}{N\bar{g}^{2}}\sum_{k}\left|g_{k}\right|^{2}\left[\tau_{k}^{-},\tau_{k}^{+}\right]\\
 & =\frac{1}{N\bar{g}^{2}}\sum_{k}\left|g_{k}\right|^{2}\left(\mathbb{I}_{k}-2\tau_{k}^{+}\tau_{k}^{-}\right)\,.
\end{align*}
Inserting the definition of $\bar{g}$ yields
\begin{equation}
\left[a,a^{\dagger}\right]=1-\frac{2}{N\bar{g}^{2}}\sum_{k}\left|g_{k}\right|^{2}\tau_{k}^{+}\tau_{k}^{-}\,.
\end{equation}
For states with only a few spins excited the second term is $\ll1$
and thus $a^{\dagger}$ and $a$ obey approximately bosonic commutation
relations,
\begin{equation}
\left[a,a^{\dagger}\right]\approx1\,.
\end{equation}
Using these collective operators the coupling term of Hamiltonian
\eqref{eq:H} becomes
\begin{equation}
\sqrt{N}\bar{g}\left(\sigma^{+}a+\sigma^{-}a^{\dagger}\right)\,.
\end{equation}
If we start with the spin ensemble in the ground state $|0\rangle$
and with the qubit excited, Rabi oscillations can transform excitations
into the ensemble that assume the form of generalized Dicke states
\begin{align}
a^{\dagger}|0\rangle & =\frac{1}{\sqrt{N}\bar{g}}\sum_{k}^{N}g_{k}\tau_{k}^{+}|0\rangle\nonumber \\
 & =\frac{1}{\sqrt{N}\bar{g}}\sum_{k}^{N}g_{k}|0_{1}\ldots1_{k}\ldots0_{N}\rangle\equiv|1\rangle\,.
\end{align}
Here $|0_{1}\ldots1_{k}\ldots0_{N}\rangle$ denotes a state where
all spins are in the ground state except the $k$-th spin. Higher
excited states are defined by
\begin{equation}
|n\rangle=\frac{1}{\sqrt{n!}}\left(a^{\dagger}\right)^{n}|0\rangle\,.
\end{equation}
If we restrict our Hilbert space to the set of states with this symmetry,
we can express $\sum_{k}^{N}\tau_{k}^{+}\tau_{k}^{-}$ in the first
part of Hamiltonian \eqref{eq:H} by collective operators as well.
To this end, we show that $\sum_{k}^{N}\tau_{k}^{+}\tau_{k}^{-}$
gives the number of collective excitations for state $|0\rangle$,

\begin{equation}
\sum_{k}^{N}\tau_{k}^{+}\tau_{k}^{-}|0\rangle=0|0\rangle
\end{equation}
and we use induction to show that if it is true for $|n\rangle$ it
is also true for $|n+1\rangle$:

\begin{align}
\sum_{k}^{N}\tau_{k}^{+}\tau_{k}^{-}|n+1\rangle & =\sum_{k}^{N}\tau_{k}^{+}\tau_{k}^{-}\frac{1}{\sqrt{n+1}}a^{\dagger}|n\rangle\label{eq:induction}\\
 & =\frac{1}{\sqrt{n+1}}\sum_{k}^{N}\tau_{k}^{+}\tau_{k}^{-}\frac{1}{\sqrt{N}\bar{g}}\sum_{l}^{N}g_{l}\tau_{l}^{+}|n\rangle\nonumber \\
 & =\frac{1}{\sqrt{n+1}}\frac{1}{\sqrt{N}\bar{g}}\Bigl(\sum_{l}^{N}g_{l}\tau_{l}^{+}\sum_{k}^{N}\tau_{k}^{+}\tau_{k}^{-}|n\rangle\nonumber \\
 & \hphantom{=\frac{1}{\sqrt{n+1}}}+\sum_{k,l}^{N}g_{l}\tau_{k}^{+} \,
 \left[\tau_{k}^{-},\tau_{l}^{+}\right]|n\rangle\Bigr)\nonumber \\
 & =n|n+1\rangle+1|n+1\rangle=\left(n+1\right)|n+1\rangle\,.\nonumber
\end{align}
In the next to last step in Eq.~(\ref{eq:induction}) we insert the relation:
$\sum_{k}^{N}\tau_{k}^{+}\tau_{k}^{-}|n\rangle = n |n\rangle$, valid by
induction.  Besides, we  use the Pauli matrices conmutation relation:
 $\left[\tau_{k}^{-},\tau_{l}^{+}\right]=\delta_{kl}\left(\mathbb{I}_{k}-2\tau_{k}^{+}\tau_{k}^{-}\right)$
 and notice that $\tau_{k}^{+}\tau_{k}^{+} = 0$.
Therefore, we can write in the subspace of the collective excitations
$|n\rangle$
\begin{equation}
\sum_{k}^{N}\tau_{k}^{+}\tau_{k}^{-}=a^{\text{\ensuremath{\dagger}}}a\,.
\end{equation}

In conclusion, restricting ourselves
to the Hilbert space of the states $|n\rangle$ the initial Hamiltonian
\eqref{eq:H} can be recast as
\begin{equation}
H=\omega_{q}\sigma^{+}\sigma^{-}+\omega_{c}a^{\dagger}a+\bar{g}\sqrt{N}\left(\sigma^{+}a+\sigma^{-}a^{\dagger}\right)\,.\label{eq:JC}
\end{equation}
Besides, in the low polarization limit $a^{\dagger}$ and $a$ obey approximately bosonic commutation relations.
Finially, It follows that the collective coupling is enhanced by a factor of
$\sqrt{N}$ as compared to the root mean square of the couplings to
the individual spins.

\section{The polaritonic basis}

The eigenstates of a single JC-Hamiltonian \eqref{eq:JC} are the
so-called polaritons and are denoted by $|n,\pm\rangle_{j}$. Here,
$n$ describes the number of excitations at a site (sum of qubit and
bosonic excitations), while the sign defines the polariton ``species''.
We now use these polaritons to express the Hamiltonian of the whole
JC-chain.

Using the polariton eigenstates $|n,\pm\rangle_{j}$ at site $j$
the individual uncoupled JC-Hamiltonians are diagonal,
\begin{equation}
H_{j}^{{\rm JC}}=\sum_{n=0}^{\infty}\sum_{\alpha=\pm}E_{n,\alpha}|n,\alpha\rangle_{j}\langle n,\alpha|_{j}\,.
\end{equation}
The individual Jaynes-Cummings energies $E_{n,\alpha}$ at a certain
site are thus given by

\begin{equation}
E_{n,\pm}=n\omega_{c}+\frac{\Delta}{2}\pm\sqrt{ng^{2}+\Delta^{2}/4}\,,\label{eq:JC energies}
\end{equation}
With the detuning $\Delta=\omega_{q}-\omega_{c}$. Note that the $|n,-\rangle$
polaritons are lower in energy than their $|n,+\rangle$ counterpart.
The eigenstates are
\begin{align}
|n,-\rangle & \equiv\cos\theta_{n}|n,{\downarrow}\rangle-\sin\theta_{n}|n-1,{\uparrow}\rangle\nonumber \\
|n,+\rangle & \equiv\sin\theta_{n}|n,{\downarrow}\rangle+\cos\theta_{n}|n-1,{\uparrow}\rangle\label{eq:JC eigenstates}
\end{align}
and $|0\rangle$ with $E_{0}=0$ where the mixing angle $\theta_{n}$
is defined as
\begin{equation}
\theta_{n}=\frac{1}{2}\arctan\left(\frac{g\sqrt{n}}{\Delta/2}\right)\ .\label{eq:mixing angle}
\end{equation}
With increasing detuning the $|n,-\rangle$ polaritons resemble more
and more pure bosonic excitations, $|n,-\rangle\approx|n\rangle|{\downarrow}\rangle$,
while the $|n,+\rangle$ polaritons exhibit an excitation in the qubit,
$|n,+\rangle\approx|n-1\rangle|{\uparrow}\rangle$.

\section{Hopping term in the polaritonic basis}

\label{app:hopping}

Next we need to express the inter-site-hopping terms,

\begin{equation}
H_{j}^{{\rm hop}}=J\left(\sigma_{j}^{+}\sigma_{j+1}+\hc\right)\,,
\end{equation}
in the local JC-basis basis as well. The operator $\sigma_{j}^{+}$
takes the form
\begin{equation}
\sigma_{j}^{+}=\sum_{n=0}^{\infty}\sum_{\alpha\beta=\pm}s_{n\alpha\beta}|n+1,\beta\rangle_{j}\langle n,\alpha|_{j}\ ,\label{eq:sigma in polaritonic basis}
\end{equation}
where the coefficients $s_{n\alpha\beta}$ are given by
\begin{equation}
\left(\begin{array}{cc}
s_{n--} & s_{n-+}\\
s_{n+-} & s_{n++}
\end{array}\right)=\left(\begin{array}{cc}
-\cos\theta_{n}\sin\theta_{n+1} & \cos\theta_{n}\cos\theta_{n+1}\\
-\sin\theta_{n}\sin\theta_{n+1} & \sin\theta_{n}\cos\theta_{n+1}
\end{array}\right)\,.
\end{equation}
These coefficients depend on the coupling $g$ and the detuning $\Delta$
between the oscillator and the qubit via~(\ref{eq:mixing angle}).
Note that the operator $\sigma_{j}^{+}$ acts on a polariton state
$|n,\alpha=\pm\rangle_{j}$ by transforming it to a linear combination
of polaritons $\pm$ with an additional excitation
\begin{equation}
\sigma_{j}^{+}|n,\alpha\rangle_{j}=\sum_{\beta=\pm}s_{n\alpha\beta}|n+1,\beta\rangle_{j}\,.
\end{equation}
The hopping term can now be rewritten in the polaritonic basis as
\begin{align}
H_{j}^{{\rm hop}}=J\sum_{nn^{\prime}} & \sum_{\alpha\alpha^{\prime}}\sum_{\beta\beta^{\prime}}s_{n\alpha\beta}|n+1,\beta\rangle_{j}\langle n,\alpha|_{j}\ \times\label{eq:coupling polaritonic basis}\\
 & s_{n^{\prime}\alpha^{\prime}\beta^{\prime}}|n^{\prime},\alpha^{\prime}\rangle_{j+1}\langle n^{\prime}+1,\beta^{\prime}|_{j+1}+\hc\nonumber
\end{align}
The hopping term therefore allows transitions between polaritons of
different types on adjacent sites.

\subsection{Effective coupling $J_{eff}$ in the case of two sites}

For a two site array initially in the state $|11\rangle\equiv|1,-\rangle_{1}\otimes|1,-\rangle_{2}$
we can derive an effective coupling constant between the two sites.
Discarding transitions between different polariton species   the dynamics
are spanned by $\{|11\rangle,|22\rangle\}$ with

\begin{equation}
|22\rangle\equiv1/\sqrt{2}\left(|2,-\rangle_{1}\otimes|0,-\rangle_{2}+|2,-\rangle_{1}\otimes|0,-\rangle_{2}\right)\,.
\end{equation}
In this case the hopping term simply reduces to

\begin{equation}
\langle11|H_{j}^{{\rm hop}}|22\rangle=J\left(s_{1--}s_{0--}+s_{1--}s_{0--}\right)\,.
\end{equation}
Therefore the effective coupling emerges as $J_{eff}=J\, s_{1--}s_{0--}$.

\subsection{Effective cavity-cavity coupling}

As in the qubit-coupled case we rewrite the harmonic oscillator operators
in the new basis
\begin{equation}
a_{j}^{+}=\sum_{n=0}^{\infty}\sum_{\alpha\beta=\pm}t_{n\alpha\beta}|n+1,\beta\rangle_{j}\langle n,\alpha|_{j}\ ,\label{eq:a in polaritonic basis}
\end{equation}
where the coefficients $t_{n\alpha\beta}$ are given by
\begin{eqnarray*}
t_{n--} & = & \cos\theta_{n}\cos\theta_{n+1}\sqrt{n+1}+\sin\theta_{n}\sin\theta_{n+1}\sqrt{n}\\
t_{n-+} & = & \cos\theta_{n}\sin\theta_{n+1}\sqrt{n+1}-\sin\theta_{n}\cos\theta_{n+1}\sqrt{n}\\
t_{n+-} & = & \sin\theta_{n}\cos\theta_{n+1}\sqrt{n+1}-\cos\theta_{n}\sin\theta_{n+1}\sqrt{n}\\
t_{n++} & = & \sin\theta_{n}\sin\theta_{n+1}\sqrt{n+1}+\cos\theta_{n}\cos\theta_{n+1}\sqrt{n}
\end{eqnarray*}
For a two site cavity coupled array the effective coupling is finally
given by $J_{eff}=J\, t_{1--}t_{0--}$.

\section{Conservation of polariton type}

\label{app:conservation}

In the previous section we have seen that, generally, the coupling
transfers polaritons from one site to another and may change the polariton
type $\left(\pm\right)$ on the sites involved.

However, not all of these possible transitions actually have to occur.
Transforming the coupling terms to the interaction picture with respect
to the uncoupled Hamiltonian; i.e.,
\begin{equation}
H_{0}=\sum_{j}H_{j}^{{\rm JC}}=\sum_{j,n,\alpha}E_{n,\alpha}|n,\alpha\rangle_{j}\langle n,\alpha|_{j}\,,
\end{equation}
we detect that the individual terms of the coupling Hamiltonian, Eq.~\eqref{eq:coupling polaritonic basis},
acquire time-dependent rotating phases
\begin{equation}
\phi_{nn^{\prime}\alpha\alpha^{\prime}\beta\beta^{\prime}}(t)=e^{it\left(E_{n+1,\beta}-E_{n,\alpha}\right)}e^{-it\left(E_{n^{\prime}+1,\beta^{\prime}}-E_{n^{\prime},\alpha^{\prime}}\right)}\,.
\end{equation}
We note that the overall frequency of the time dependent phase --
given in terms of the energies of the states it is coupled -- broadly
varies in magnitude. Thus, under the premise of a small inter-site
coupling $J$, we can identify fast oscillating terms and neglect
those within a rotating wave approximation (RWA). Particularly, terms
connecting states which are far apart in energy ($\gg J$) and thus
are rotating fast, are the ones that can be neglected.

This treatment is certainly useful in the case of large detuning or
strong coupling strength $g$. Then, polaritons of different sign lie far
apart in energy. Thus, their interconversion can  safely be neglected
and the initial polariton species ($\pm$) is thus conserved.\bibliographystyle{apsrev4-1}

\begin{thebibliography}{38}%
\makeatletter
\providecommand \@ifxundefined [1]{%
 \@ifx{#1\undefined}
}%
\providecommand \@ifnum [1]{%
 \ifnum #1\expandafter \@firstoftwo
 \else \expandafter \@secondoftwo
 \fi
}%
\providecommand \@ifx [1]{%
 \ifx #1\expandafter \@firstoftwo
 \else \expandafter \@secondoftwo
 \fi
}%
\providecommand \natexlab [1]{#1}%
\providecommand \enquote  [1]{``#1''}%
\providecommand \bibnamefont  [1]{#1}%
\providecommand \bibfnamefont [1]{#1}%
\providecommand \citenamefont [1]{#1}%
\providecommand \href@noop [0]{\@secondoftwo}%
\providecommand \href [0]{\begingroup \@sanitize@url \@href}%
\providecommand \@href[1]{\@@startlink{#1}\@@href}%
\providecommand \@@href[1]{\endgroup#1\@@endlink}%
\providecommand \@sanitize@url [0]{\catcode `\\12\catcode `\$12\catcode
  `\&12\catcode `\#12\catcode `\^12\catcode `\_12\catcode `\%12\relax}%
\providecommand \@@startlink[1]{}%
\providecommand \@@endlink[0]{}%
\providecommand \url  [0]{\begingroup\@sanitize@url \@url }%
\providecommand \@url [1]{\endgroup\@href {#1}{\urlprefix }}%
\providecommand \urlprefix  [0]{URL }%
\providecommand \Eprint [0]{\href }%
\providecommand \doibase [0]{http://dx.doi.org/}%
\providecommand \selectlanguage [0]{\@gobble}%
\providecommand \bibinfo  [0]{\@secondoftwo}%
\providecommand \bibfield  [0]{\@secondoftwo}%
\providecommand \translation [1]{[#1]}%
\providecommand \BibitemOpen [0]{}%
\providecommand \bibitemStop [0]{}%
\providecommand \bibitemNoStop [0]{.\EOS\space}%
\providecommand \EOS [0]{\spacefactor3000\relax}%
\providecommand \BibitemShut  [1]{\csname bibitem#1\endcsname}%
\let\auto@bib@innerbib\@empty
\bibitem [{\citenamefont {You}\ and\ \citenamefont
  {Nori}(2011)}]{you2011atomicphysics}%
  \BibitemOpen
  \bibfield  {author} {\bibinfo {author} {\bibfnamefont {J.~Q.}\ \bibnamefont
  {You}}\ and\ \bibinfo {author} {\bibfnamefont {F.}~\bibnamefont {Nori}},\
  }\href@noop {} {\bibfield  {journal} {\bibinfo  {journal} {Nature (London)}\
  }\textbf {\bibinfo {volume} {474}},\ \bibinfo {pages} {589} (\bibinfo {year}
  {2011})}\BibitemShut {NoStop}%
\bibitem [{\citenamefont {Buluta}\ \emph {et~al.}(2011)\citenamefont {Buluta},
  \citenamefont {Ashhab},\ and\ \citenamefont {Nori}}]{buluta_natural_2011}%
  \BibitemOpen
  \bibfield  {author} {\bibinfo {author} {\bibfnamefont {I.}~\bibnamefont
  {Buluta}}, \bibinfo {author} {\bibfnamefont {S.}~\bibnamefont {Ashhab}}, \
  and\ \bibinfo {author} {\bibfnamefont {F.}~\bibnamefont {Nori}},\ }\href@noop
  {} {\bibfield  {journal} {\bibinfo  {journal} {Rep. Prog. Phys.}\ }\textbf
  {\bibinfo {volume} {74}},\ \bibinfo {pages} {104401} (\bibinfo {year}
  {2011})}\BibitemShut {NoStop}%
\bibitem [{\citenamefont {Ladd}\ \emph {et~al.}(2010)\citenamefont {Ladd},
  \citenamefont {Jelezko}, \citenamefont {Laflamme}, \citenamefont {Nakamura},
  \citenamefont {Monroe},\ and\ \citenamefont {O'Brien}}]{ladd2010quantum}%
  \BibitemOpen
  \bibfield  {author} {\bibinfo {author} {\bibfnamefont {T.~D.}\ \bibnamefont
  {Ladd}}, \bibinfo {author} {\bibfnamefont {F.}~\bibnamefont {Jelezko}},
  \bibinfo {author} {\bibfnamefont {R.}~\bibnamefont {Laflamme}}, \bibinfo
  {author} {\bibfnamefont {Y.}~\bibnamefont {Nakamura}}, \bibinfo {author}
  {\bibfnamefont {C.}~\bibnamefont {Monroe}}, \ and\ \bibinfo {author}
  {\bibfnamefont {J.~L.}\ \bibnamefont {O'Brien}},\ }\href@noop {} {\bibfield
  {journal} {\bibinfo  {journal} {Nature (London)}\ }\textbf {\bibinfo {volume}
  {464}},\ \bibinfo {pages} {45} (\bibinfo {year} {2010})}\BibitemShut
  {NoStop}%
\bibitem [{\citenamefont {Feynman}(1982)}]{feynman1982simulating}%
  \BibitemOpen
  \bibfield  {author} {\bibinfo {author} {\bibfnamefont {R.~P.}\ \bibnamefont
  {Feynman}},\ }\href@noop {} {\bibfield  {journal} {\bibinfo  {journal}
  {International Journal of Theoretical Physics}\ }\textbf {\bibinfo {volume}
  {21}},\ \bibinfo {pages} {467} (\bibinfo {year} {1982})}\BibitemShut
  {NoStop}%
\bibitem [{\citenamefont {Houck}\ \emph {et~al.}(2012)\citenamefont {Houck},
  \citenamefont {T\"{u}reci},\ and\ \citenamefont
  {Koch}}]{houck2012onchipquantum}%
  \BibitemOpen
  \bibfield  {author} {\bibinfo {author} {\bibfnamefont {A.~A.}\ \bibnamefont
  {Houck}}, \bibinfo {author} {\bibfnamefont {H.~E.}\ \bibnamefont
  {T\"{u}reci}}, \ and\ \bibinfo {author} {\bibfnamefont {J.}~\bibnamefont
  {Koch}},\ }\href@noop {} {\bibfield  {journal} {\bibinfo  {journal} {Nat.
  Phys.}\ }\textbf {\bibinfo {volume} {8}},\ \bibinfo {pages} {292} (\bibinfo
  {year} {2012})}\BibitemShut {NoStop}%
\bibitem [{\citenamefont {Ams\"{u}ss}\ \emph {et~al.}(2011)\citenamefont
  {Ams\"{u}ss}, \citenamefont {Koller}, \citenamefont {N\"{o}bauer},
  \citenamefont {Putz}, \citenamefont {Rotter}, \citenamefont {Sandner},
  \citenamefont {Schneider}, \citenamefont {Schramb\"{o}ck}, \citenamefont
  {Steinhauser}, \citenamefont {Ritsch}, \citenamefont {Schmiedmayer},\ and\
  \citenamefont {Majer}}]{amsuss2011cavityqed}%
  \BibitemOpen
  \bibfield  {author} {\bibinfo {author} {\bibfnamefont {R.}~\bibnamefont
  {Ams\"{u}ss}}, \bibinfo {author} {\bibfnamefont {C.}~\bibnamefont {Koller}},
  \bibinfo {author} {\bibfnamefont {T.}~\bibnamefont {N\"{o}bauer}}, \bibinfo
  {author} {\bibfnamefont {S.}~\bibnamefont {Putz}}, \bibinfo {author}
  {\bibfnamefont {S.}~\bibnamefont {Rotter}}, \bibinfo {author} {\bibfnamefont
  {K.}~\bibnamefont {Sandner}}, \bibinfo {author} {\bibfnamefont
  {S.}~\bibnamefont {Schneider}}, \bibinfo {author} {\bibfnamefont
  {M.}~\bibnamefont {Schramb\"{o}ck}}, \bibinfo {author} {\bibfnamefont
  {G.}~\bibnamefont {Steinhauser}}, \bibinfo {author} {\bibfnamefont
  {H.}~\bibnamefont {Ritsch}}, \bibinfo {author} {\bibfnamefont
  {J.}~\bibnamefont {Schmiedmayer}}, \ and\ \bibinfo {author} {\bibfnamefont
  {J.}~\bibnamefont {Majer}},\ }\href@noop {} {\bibfield  {journal} {\bibinfo
  {journal} {Phys. Rev. Lett.}\ }\textbf {\bibinfo {volume} {107}},\ \bibinfo
  {pages} {060502} (\bibinfo {year} {2011})}\BibitemShut {NoStop}%
\bibitem [{\citenamefont {Kubo}\ \emph {et~al.}(2011)\citenamefont {Kubo},
  \citenamefont {Grezes}, \citenamefont {Dewes}, \citenamefont {Umeda},
  \citenamefont {Isoya}, \citenamefont {Sumiya}, \citenamefont {Morishita},
  \citenamefont {Abe}, \citenamefont {Onoda}, \citenamefont {Ohshima},
  \citenamefont {Jacques}, \citenamefont {Dr\'{e}au}, \citenamefont {Roch},
  \citenamefont {Diniz}, \citenamefont {Auffeves}, \citenamefont {Vion},
  \citenamefont {Esteve},\ and\ \citenamefont
  {Bertet}}]{kubo2011hybridquantum}%
  \BibitemOpen
  \bibfield  {author} {\bibinfo {author} {\bibfnamefont {Y.}~\bibnamefont
  {Kubo}}, \bibinfo {author} {\bibfnamefont {C.}~\bibnamefont {Grezes}},
  \bibinfo {author} {\bibfnamefont {A.}~\bibnamefont {Dewes}}, \bibinfo
  {author} {\bibfnamefont {T.}~\bibnamefont {Umeda}}, \bibinfo {author}
  {\bibfnamefont {J.}~\bibnamefont {Isoya}}, \bibinfo {author} {\bibfnamefont
  {H.}~\bibnamefont {Sumiya}}, \bibinfo {author} {\bibfnamefont
  {N.}~\bibnamefont {Morishita}}, \bibinfo {author} {\bibfnamefont
  {H.}~\bibnamefont {Abe}}, \bibinfo {author} {\bibfnamefont {S.}~\bibnamefont
  {Onoda}}, \bibinfo {author} {\bibfnamefont {T.}~\bibnamefont {Ohshima}},
  \bibinfo {author} {\bibfnamefont {V.}~\bibnamefont {Jacques}}, \bibinfo
  {author} {\bibfnamefont {A.}~\bibnamefont {Dr\'{e}au}}, \bibinfo {author}
  {\bibfnamefont {J.~F.}\ \bibnamefont {Roch}}, \bibinfo {author}
  {\bibfnamefont {I.}~\bibnamefont {Diniz}}, \bibinfo {author} {\bibfnamefont
  {A.}~\bibnamefont {Auffeves}}, \bibinfo {author} {\bibfnamefont
  {D.}~\bibnamefont {Vion}}, \bibinfo {author} {\bibfnamefont {D.}~\bibnamefont
  {Esteve}}, \ and\ \bibinfo {author} {\bibfnamefont {P.}~\bibnamefont
  {Bertet}},\ }\href@noop {} {\bibfield  {journal} {\bibinfo  {journal} {Phys.
  Rev. Lett.}\ }\textbf {\bibinfo {volume} {107}},\ \bibinfo {pages} {220501}
  (\bibinfo {year} {2011})}\BibitemShut {NoStop}%
\bibitem [{\citenamefont {Kubo}\ \emph {et~al.}(2010)\citenamefont {Kubo},
  \citenamefont {Ong}, \citenamefont {Bertet}, \citenamefont {Vion},
  \citenamefont {Jacques}, \citenamefont {Zheng}, \citenamefont {Dr\'{e}au},
  \citenamefont {Roch}, \citenamefont {Auffeves}, \citenamefont {Jelezko},
  \citenamefont {Wrachtrup}, \citenamefont {Barthe}, \citenamefont {Bergonzo},\
  and\ \citenamefont {Esteve}}]{kubo2010strongcoupling}%
  \BibitemOpen
  \bibfield  {author} {\bibinfo {author} {\bibfnamefont {Y.}~\bibnamefont
  {Kubo}}, \bibinfo {author} {\bibfnamefont {F.~R.}\ \bibnamefont {Ong}},
  \bibinfo {author} {\bibfnamefont {P.}~\bibnamefont {Bertet}}, \bibinfo
  {author} {\bibfnamefont {D.}~\bibnamefont {Vion}}, \bibinfo {author}
  {\bibfnamefont {V.}~\bibnamefont {Jacques}}, \bibinfo {author} {\bibfnamefont
  {D.}~\bibnamefont {Zheng}}, \bibinfo {author} {\bibfnamefont
  {A.}~\bibnamefont {Dr\'{e}au}}, \bibinfo {author} {\bibfnamefont {J.~F.}\
  \bibnamefont {Roch}}, \bibinfo {author} {\bibfnamefont {A.}~\bibnamefont
  {Auffeves}}, \bibinfo {author} {\bibfnamefont {F.}~\bibnamefont {Jelezko}},
  \bibinfo {author} {\bibfnamefont {J.}~\bibnamefont {Wrachtrup}}, \bibinfo
  {author} {\bibfnamefont {M.~F.}\ \bibnamefont {Barthe}}, \bibinfo {author}
  {\bibfnamefont {P.}~\bibnamefont {Bergonzo}}, \ and\ \bibinfo {author}
  {\bibfnamefont {D.}~\bibnamefont {Esteve}},\ }\href@noop {} {\bibfield
  {journal} {\bibinfo  {journal} {Phys. Rev. Lett.}\ }\textbf {\bibinfo
  {volume} {105}},\ \bibinfo {pages} {140502} (\bibinfo {year}
  {2010})}\BibitemShut {NoStop}%
\bibitem [{\citenamefont {Schuster}\ \emph {et~al.}(2010)\citenamefont
  {Schuster}, \citenamefont {Sears}, \citenamefont {Ginossar}, \citenamefont
  {{DiCarlo}}, \citenamefont {Frunzio}, \citenamefont {Morton}, \citenamefont
  {Wu}, \citenamefont {Briggs}, \citenamefont {Buckley}, \citenamefont
  {Awschalom},\ and\ \citenamefont
  {Schoelkopf}}]{schuster2010highcooperativity}%
  \BibitemOpen
  \bibfield  {author} {\bibinfo {author} {\bibfnamefont {D.~I.}\ \bibnamefont
  {Schuster}}, \bibinfo {author} {\bibfnamefont {A.~P.}\ \bibnamefont {Sears}},
  \bibinfo {author} {\bibfnamefont {E.}~\bibnamefont {Ginossar}}, \bibinfo
  {author} {\bibfnamefont {L.}~\bibnamefont {{DiCarlo}}}, \bibinfo {author}
  {\bibfnamefont {L.}~\bibnamefont {Frunzio}}, \bibinfo {author} {\bibfnamefont
  {J.~J.~L.}\ \bibnamefont {Morton}}, \bibinfo {author} {\bibfnamefont
  {H.}~\bibnamefont {Wu}}, \bibinfo {author} {\bibfnamefont {G.~A.~D.}\
  \bibnamefont {Briggs}}, \bibinfo {author} {\bibfnamefont {B.~B.}\
  \bibnamefont {Buckley}}, \bibinfo {author} {\bibfnamefont {D.~D.}\
  \bibnamefont {Awschalom}}, \ and\ \bibinfo {author} {\bibfnamefont {R.~J.}\
  \bibnamefont {Schoelkopf}},\ }\href@noop {} {\bibfield  {journal} {\bibinfo
  {journal} {Phys. Rev. Lett.}\ }\textbf {\bibinfo {volume} {105}},\ \bibinfo
  {pages} {140501} (\bibinfo {year} {2010})}\BibitemShut {NoStop}%
\bibitem [{\citenamefont {Imamo\u{g}lu}(2009)}]{imamoglu2009cavityqed}%
  \BibitemOpen
  \bibfield  {author} {\bibinfo {author} {\bibfnamefont {A.}~\bibnamefont
  {Imamo\u{g}lu}},\ }\href@noop {} {\bibfield  {journal} {\bibinfo  {journal}
  {Phys. Rev. Lett.}\ }\textbf {\bibinfo {volume} {102}},\ \bibinfo {pages}
  {083602} (\bibinfo {year} {2009})}\BibitemShut {NoStop}%
\bibitem [{\citenamefont {Rabl}\ \emph {et~al.}(2006)\citenamefont {Rabl},
  \citenamefont {{DeMille}}, \citenamefont {Doyle}, \citenamefont {Lukin},
  \citenamefont {Schoelkopf},\ and\ \citenamefont
  {Zoller}}]{rabl2006hybridquantum}%
  \BibitemOpen
  \bibfield  {author} {\bibinfo {author} {\bibfnamefont {P.}~\bibnamefont
  {Rabl}}, \bibinfo {author} {\bibfnamefont {D.}~\bibnamefont {{DeMille}}},
  \bibinfo {author} {\bibfnamefont {J.~M.}\ \bibnamefont {Doyle}}, \bibinfo
  {author} {\bibfnamefont {M.~D.}\ \bibnamefont {Lukin}}, \bibinfo {author}
  {\bibfnamefont {R.~J.}\ \bibnamefont {Schoelkopf}}, \ and\ \bibinfo {author}
  {\bibfnamefont {P.}~\bibnamefont {Zoller}},\ }\href@noop {} {\bibfield
  {journal} {\bibinfo  {journal} {Phys. Rev. Lett.}\ }\textbf {\bibinfo
  {volume} {97}},\ \bibinfo {pages} {033003} (\bibinfo {year}
  {2006})}\BibitemShut {NoStop}%
\bibitem [{\citenamefont {Verdu}\ \emph {et~al.}(2009)\citenamefont {Verdu},
  \citenamefont {Zoubi}, \citenamefont {Koller}, \citenamefont {Majer},
  \citenamefont {Ritsch},\ and\ \citenamefont
  {Schmiedmayer}}]{verdu2009strongmagnetic}%
  \BibitemOpen
  \bibfield  {author} {\bibinfo {author} {\bibfnamefont {J.}~\bibnamefont
  {Verdu}}, \bibinfo {author} {\bibfnamefont {H.}~\bibnamefont {Zoubi}},
  \bibinfo {author} {\bibfnamefont {C.}~\bibnamefont {Koller}}, \bibinfo
  {author} {\bibfnamefont {J.}~\bibnamefont {Majer}}, \bibinfo {author}
  {\bibfnamefont {H.}~\bibnamefont {Ritsch}}, \ and\ \bibinfo {author}
  {\bibfnamefont {J.}~\bibnamefont {Schmiedmayer}},\ }\href@noop {} {\bibfield
  {journal} {\bibinfo  {journal} {Phys. Rev. Lett.}\ }\textbf {\bibinfo
  {volume} {103}},\ \bibinfo {pages} {043603} (\bibinfo {year}
  {2009})}\BibitemShut {NoStop}%
\bibitem [{\citenamefont {Wesenberg}\ \emph {et~al.}(2009)\citenamefont
  {Wesenberg}, \citenamefont {Ardavan}, \citenamefont {Briggs}, \citenamefont
  {Morton}, \citenamefont {Schoelkopf}, \citenamefont {Schuster},\ and\
  \citenamefont {Molmer}}]{wesenberg2009quantum}%
  \BibitemOpen
  \bibfield  {author} {\bibinfo {author} {\bibfnamefont {J.~H.}\ \bibnamefont
  {Wesenberg}}, \bibinfo {author} {\bibfnamefont {A.}~\bibnamefont {Ardavan}},
  \bibinfo {author} {\bibfnamefont {G.~A.~D.}\ \bibnamefont {Briggs}}, \bibinfo
  {author} {\bibfnamefont {J.~J.~L.}\ \bibnamefont {Morton}}, \bibinfo {author}
  {\bibfnamefont {R.~J.}\ \bibnamefont {Schoelkopf}}, \bibinfo {author}
  {\bibfnamefont {D.~I.}\ \bibnamefont {Schuster}}, \ and\ \bibinfo {author}
  {\bibfnamefont {K.}~\bibnamefont {Molmer}},\ }\href@noop {} {\bibfield
  {journal} {\bibinfo  {journal} {Phys. Rev. Lett.}\ }\textbf {\bibinfo
  {volume} {103}},\ \bibinfo {pages} {070502} (\bibinfo {year}
  {2009})}\BibitemShut {NoStop}%
\bibitem [{\citenamefont {Xiang}\ \emph {et~al.}(2012)\citenamefont {Xiang},
  \citenamefont {Ashhab}, \citenamefont {You},\ and\ \citenamefont
  {Nori}}]{xiang2012hybridquantum}%
  \BibitemOpen
  \bibfield  {author} {\bibinfo {author} {\bibfnamefont {Z.}~\bibnamefont
  {Xiang}}, \bibinfo {author} {\bibfnamefont {S.}~\bibnamefont {Ashhab}},
  \bibinfo {author} {\bibfnamefont {J.~Q.}\ \bibnamefont {You}}, \ and\
  \bibinfo {author} {\bibfnamefont {F.}~\bibnamefont {Nori}},\ }\href
  {http://arxiv.org/abs/1204.2137} {\bibfield  {journal} {\bibinfo  {journal}
  {{arXiv:1204.2137}}\ } (\bibinfo {year} {2012})}\BibitemShut {NoStop}%
\bibitem [{\citenamefont {Hartmann}\ \emph {et~al.}(2006)\citenamefont
  {Hartmann}, \citenamefont {Brandao},\ and\ \citenamefont
  {Plenio}}]{hartmann2006strongly}%
  \BibitemOpen
  \bibfield  {author} {\bibinfo {author} {\bibfnamefont {M.~J.}\ \bibnamefont
  {Hartmann}}, \bibinfo {author} {\bibfnamefont {F.~G.}\ \bibnamefont
  {Brandao}}, \ and\ \bibinfo {author} {\bibfnamefont {M.~B.}\ \bibnamefont
  {Plenio}},\ }\href@noop {} {\bibfield  {journal} {\bibinfo  {journal} {Nat.
  Phys.}\ }\textbf {\bibinfo {volume} {2}},\ \bibinfo {pages}
  {849{\textendash}855} (\bibinfo {year} {2006})}\BibitemShut {NoStop}%
\bibitem [{\citenamefont {Greentree}\ \emph {et~al.}(2006)\citenamefont
  {Greentree}, \citenamefont {Tahan}, \citenamefont {Cole},\ and\ \citenamefont
  {Hollenberg}}]{greentree2006quantum}%
  \BibitemOpen
  \bibfield  {author} {\bibinfo {author} {\bibfnamefont {A.~D.}\ \bibnamefont
  {Greentree}}, \bibinfo {author} {\bibfnamefont {C.}~\bibnamefont {Tahan}},
  \bibinfo {author} {\bibfnamefont {J.~H.}\ \bibnamefont {Cole}}, \ and\
  \bibinfo {author} {\bibfnamefont {L.~C.~L.}\ \bibnamefont {Hollenberg}},\
  }\href@noop {} {\bibfield  {journal} {\bibinfo  {journal} {Nat. Phys.}\
  }\textbf {\bibinfo {volume} {2}},\ \bibinfo {pages} {856} (\bibinfo {year}
  {2006})}\BibitemShut {NoStop}%
\bibitem [{\citenamefont {Angelakis}\ \emph {et~al.}(2007)\citenamefont
  {Angelakis}, \citenamefont {Santos},\ and\ \citenamefont
  {Bose}}]{angelakis2007photonblockadeinduced}%
  \BibitemOpen
  \bibfield  {author} {\bibinfo {author} {\bibfnamefont {D.~G.}\ \bibnamefont
  {Angelakis}}, \bibinfo {author} {\bibfnamefont {M.~F.}\ \bibnamefont
  {Santos}}, \ and\ \bibinfo {author} {\bibfnamefont {S.}~\bibnamefont
  {Bose}},\ }\href@noop {} {\bibfield  {journal} {\bibinfo  {journal} {Phys.
  Rev. A}\ }\textbf {\bibinfo {volume} {76}},\ \bibinfo {pages} {031805}
  (\bibinfo {year} {2007})}\BibitemShut {NoStop}%
\bibitem [{\citenamefont {Hartmann}\ \emph {et~al.}(2008)\citenamefont
  {Hartmann}, \citenamefont {Brandao},\ and\ \citenamefont
  {Plenio}}]{hartmann2008quantum}%
  \BibitemOpen
  \bibfield  {author} {\bibinfo {author} {\bibfnamefont {M.~J.}\ \bibnamefont
  {Hartmann}}, \bibinfo {author} {\bibfnamefont {F.~G.}\ \bibnamefont
  {Brandao}}, \ and\ \bibinfo {author} {\bibfnamefont {M.~B.}\ \bibnamefont
  {Plenio}},\ }\href@noop {} {\bibfield  {journal} {\bibinfo  {journal} {Laser
  \& Photon. Rev.}\ }\textbf {\bibinfo {volume} {2}},\ \bibinfo {pages}
  {527{\textendash}556} (\bibinfo {year} {2008})}\BibitemShut {NoStop}%
\bibitem [{\citenamefont {Marcos}\ \emph {et~al.}(2010)\citenamefont {Marcos},
  \citenamefont {Wubs}, \citenamefont {Taylor}, \citenamefont {Aguado},
  \citenamefont {Lukin},\ and\ \citenamefont {Sorensen}}]{marcos2010coupling}%
  \BibitemOpen
  \bibfield  {author} {\bibinfo {author} {\bibfnamefont {D.}~\bibnamefont
  {Marcos}}, \bibinfo {author} {\bibfnamefont {M.}~\bibnamefont {Wubs}},
  \bibinfo {author} {\bibfnamefont {J.~M.}\ \bibnamefont {Taylor}}, \bibinfo
  {author} {\bibfnamefont {R.}~\bibnamefont {Aguado}}, \bibinfo {author}
  {\bibfnamefont {M.~D.}\ \bibnamefont {Lukin}}, \ and\ \bibinfo {author}
  {\bibfnamefont {A.~S.}\ \bibnamefont {Sorensen}},\ }\href@noop {} {\bibfield
  {journal} {\bibinfo  {journal} {Phys. Rev. Lett.}\ }\textbf {\bibinfo
  {volume} {105}},\ \bibinfo {pages} {210501} (\bibinfo {year}
  {2010})}\BibitemShut {NoStop}%
\bibitem [{\citenamefont {Twamley}\ and\ \citenamefont
  {Barrett}(2010)}]{twamley2010superconducting}%
  \BibitemOpen
  \bibfield  {author} {\bibinfo {author} {\bibfnamefont {J.}~\bibnamefont
  {Twamley}}\ and\ \bibinfo {author} {\bibfnamefont {S.~D.}\ \bibnamefont
  {Barrett}},\ }\href@noop {} {\bibfield  {journal} {\bibinfo  {journal} {Phys.
  Rev. B}\ }\textbf {\bibinfo {volume} {81}},\ \bibinfo {pages} {241202}
  (\bibinfo {year} {2010})}\BibitemShut {NoStop}%
\bibitem [{\citenamefont {Zhu}\ \emph {et~al.}(2011)\citenamefont {Zhu},
  \citenamefont {Saito}, \citenamefont {Kemp}, \citenamefont {Kakuyanagi},
  \citenamefont {Karimoto}, \citenamefont {Nakano}, \citenamefont {Munro},
  \citenamefont {Tokura}, \citenamefont {Everitt}, \citenamefont {Nemoto},
  \citenamefont {Kasu}, \citenamefont {Mizuochi},\ and\ \citenamefont
  {Semba}}]{zhu2011coherent}%
  \BibitemOpen
  \bibfield  {author} {\bibinfo {author} {\bibfnamefont {X.}~\bibnamefont
  {Zhu}}, \bibinfo {author} {\bibfnamefont {S.}~\bibnamefont {Saito}}, \bibinfo
  {author} {\bibfnamefont {A.}~\bibnamefont {Kemp}}, \bibinfo {author}
  {\bibfnamefont {K.}~\bibnamefont {Kakuyanagi}}, \bibinfo {author}
  {\bibfnamefont {S.-i.}\ \bibnamefont {Karimoto}}, \bibinfo {author}
  {\bibfnamefont {H.}~\bibnamefont {Nakano}}, \bibinfo {author} {\bibfnamefont
  {W.~J.}\ \bibnamefont {Munro}}, \bibinfo {author} {\bibfnamefont
  {Y.}~\bibnamefont {Tokura}}, \bibinfo {author} {\bibfnamefont {M.~S.}\
  \bibnamefont {Everitt}}, \bibinfo {author} {\bibfnamefont {K.}~\bibnamefont
  {Nemoto}}, \bibinfo {author} {\bibfnamefont {M.}~\bibnamefont {Kasu}},
  \bibinfo {author} {\bibfnamefont {N.}~\bibnamefont {Mizuochi}}, \ and\
  \bibinfo {author} {\bibfnamefont {K.}~\bibnamefont {Semba}},\ }\href@noop {}
  {\bibfield  {journal} {\bibinfo  {journal} {Nature (London)}\ }\textbf
  {\bibinfo {volume} {478}},\ \bibinfo {pages} {221} (\bibinfo {year}
  {2011})}\BibitemShut {NoStop}%
\bibitem [{\citenamefont {Harrabi}\ \emph {et~al.}(2009)\citenamefont
  {Harrabi}, \citenamefont {Yoshihara}, \citenamefont {Niskanen}, \citenamefont
  {Nakamura},\ and\ \citenamefont {Tsai}}]{harrabi2009engineered}%
  \BibitemOpen
  \bibfield  {author} {\bibinfo {author} {\bibfnamefont {K.}~\bibnamefont
  {Harrabi}}, \bibinfo {author} {\bibfnamefont {F.}~\bibnamefont {Yoshihara}},
  \bibinfo {author} {\bibfnamefont {A.~O.}\ \bibnamefont {Niskanen}}, \bibinfo
  {author} {\bibfnamefont {Y.}~\bibnamefont {Nakamura}}, \ and\ \bibinfo
  {author} {\bibfnamefont {J.~S.}\ \bibnamefont {Tsai}},\ }\href@noop {}
  {\bibfield  {journal} {\bibinfo  {journal} {Phys. Rev. B}\ }\textbf {\bibinfo
  {volume} {79}},\ \bibinfo {pages} {020507} (\bibinfo {year}
  {2009})}\BibitemShut {NoStop}%
\bibitem [{\citenamefont {Hime}\ \emph {et~al.}(2006)\citenamefont {Hime},
  \citenamefont {Reichardt}, \citenamefont {Plourde}, \citenamefont
  {Robertson}, \citenamefont {Wu}, \citenamefont {Ustinov},\ and\ \citenamefont
  {Clarke}}]{hime2006solidstate}%
  \BibitemOpen
  \bibfield  {author} {\bibinfo {author} {\bibfnamefont {T.}~\bibnamefont
  {Hime}}, \bibinfo {author} {\bibfnamefont {P.~A.}\ \bibnamefont {Reichardt}},
  \bibinfo {author} {\bibfnamefont {B.~L.~T.}\ \bibnamefont {Plourde}},
  \bibinfo {author} {\bibfnamefont {T.~L.}\ \bibnamefont {Robertson}}, \bibinfo
  {author} {\bibfnamefont {C.}~\bibnamefont {Wu}}, \bibinfo {author}
  {\bibfnamefont {A.~V.}\ \bibnamefont {Ustinov}}, \ and\ \bibinfo {author}
  {\bibfnamefont {J.}~\bibnamefont {Clarke}},\ }\href@noop {} {\bibfield
  {journal} {\bibinfo  {journal} {Science}\ }\textbf {\bibinfo {volume}
  {314}},\ \bibinfo {pages} {1427 } (\bibinfo {year} {2006})}\BibitemShut
  {NoStop}%
\bibitem [{\citenamefont {Plourde}\ \emph {et~al.}(2004)\citenamefont
  {Plourde}, \citenamefont {Zhang}, \citenamefont {Whaley}, \citenamefont
  {Wilhelm}, \citenamefont {Robertson}, \citenamefont {Hime}, \citenamefont
  {Linzen}, \citenamefont {Reichardt}, \citenamefont {Wu},\ and\ \citenamefont
  {Clarke}}]{plourde2004entangling}%
  \BibitemOpen
  \bibfield  {author} {\bibinfo {author} {\bibfnamefont {B.~L.~T.}\
  \bibnamefont {Plourde}}, \bibinfo {author} {\bibfnamefont {J.}~\bibnamefont
  {Zhang}}, \bibinfo {author} {\bibfnamefont {K.~B.}\ \bibnamefont {Whaley}},
  \bibinfo {author} {\bibfnamefont {F.~K.}\ \bibnamefont {Wilhelm}}, \bibinfo
  {author} {\bibfnamefont {T.~L.}\ \bibnamefont {Robertson}}, \bibinfo {author}
  {\bibfnamefont {T.}~\bibnamefont {Hime}}, \bibinfo {author} {\bibfnamefont
  {S.}~\bibnamefont {Linzen}}, \bibinfo {author} {\bibfnamefont {P.~A.}\
  \bibnamefont {Reichardt}}, \bibinfo {author} {\bibfnamefont {C.~E.}\
  \bibnamefont {Wu}}, \ and\ \bibinfo {author} {\bibfnamefont {J.}~\bibnamefont
  {Clarke}},\ }\href@noop {} {\bibfield  {journal} {\bibinfo  {journal} {Phys.
  Rev. B}\ }\textbf {\bibinfo {volume} {70}},\ \bibinfo {pages} {140501}
  (\bibinfo {year} {2004})}\BibitemShut {NoStop}%
\bibitem [{\citenamefont {Niskanen}\ \emph {et~al.}(2007)\citenamefont
  {Niskanen}, \citenamefont {Harrabi}, \citenamefont {Yoshihara}, \citenamefont
  {Nakamura}, \citenamefont {Lloyd},\ and\ \citenamefont
  {Tsai}}]{niskanen2007quantum}%
  \BibitemOpen
  \bibfield  {author} {\bibinfo {author} {\bibfnamefont {A.~O.}\ \bibnamefont
  {Niskanen}}, \bibinfo {author} {\bibfnamefont {K.}~\bibnamefont {Harrabi}},
  \bibinfo {author} {\bibfnamefont {F.}~\bibnamefont {Yoshihara}}, \bibinfo
  {author} {\bibfnamefont {Y.}~\bibnamefont {Nakamura}}, \bibinfo {author}
  {\bibfnamefont {S.}~\bibnamefont {Lloyd}}, \ and\ \bibinfo {author}
  {\bibfnamefont {J.~S.}\ \bibnamefont {Tsai}},\ }\href@noop {} {\bibfield
  {journal} {\bibinfo  {journal} {Science}\ }\textbf {\bibinfo {volume}
  {316}},\ \bibinfo {pages} {723} (\bibinfo {year} {2007})}\BibitemShut
  {NoStop}%
\bibitem [{\citenamefont {van~der Ploeg}\ \emph {et~al.}(2007)\citenamefont
  {van~der Ploeg}, \citenamefont {Izmalkov}, \citenamefont {van~den Brink},
  \citenamefont {H\"{u}bner}, \citenamefont {Grajcar}, \citenamefont {Ilichev},
  \citenamefont {Meyer},\ and\ \citenamefont
  {Zagoskin}}]{vanderploeg2007controllable}%
  \BibitemOpen
  \bibfield  {author} {\bibinfo {author} {\bibfnamefont {S.~H.~W.}\
  \bibnamefont {van~der Ploeg}}, \bibinfo {author} {\bibfnamefont
  {A.}~\bibnamefont {Izmalkov}}, \bibinfo {author} {\bibfnamefont {A.~M.}\
  \bibnamefont {van~den Brink}}, \bibinfo {author} {\bibfnamefont
  {U.}~\bibnamefont {H\"{u}bner}}, \bibinfo {author} {\bibfnamefont
  {M.}~\bibnamefont {Grajcar}}, \bibinfo {author} {\bibfnamefont
  {E.}~\bibnamefont {Ilichev}}, \bibinfo {author} {\bibfnamefont {H.~G.}\
  \bibnamefont {Meyer}}, \ and\ \bibinfo {author} {\bibfnamefont {A.~M.}\
  \bibnamefont {Zagoskin}},\ }\href@noop {} {\bibfield  {journal} {\bibinfo
  {journal} {Phys. Rev. Lett.}\ }\textbf {\bibinfo {volume} {98}},\ \bibinfo
  {pages} {057004} (\bibinfo {year} {2007})}\BibitemShut {NoStop}%
\bibitem [{\citenamefont {Harris}\ \emph {et~al.}(2009)\citenamefont {Harris},
  \citenamefont {Lanting}, \citenamefont {Berkley}, \citenamefont {Johansson},
  \citenamefont {Johnson}, \citenamefont {Bunyk}, \citenamefont {Ladizinsky},
  \citenamefont {Ladizinsky}, \citenamefont {Oh},\ and\ \citenamefont
  {Han}}]{harris2009compound}%
  \BibitemOpen
  \bibfield  {author} {\bibinfo {author} {\bibfnamefont {R.}~\bibnamefont
  {Harris}}, \bibinfo {author} {\bibfnamefont {T.}~\bibnamefont {Lanting}},
  \bibinfo {author} {\bibfnamefont {A.~J.}\ \bibnamefont {Berkley}}, \bibinfo
  {author} {\bibfnamefont {J.}~\bibnamefont {Johansson}}, \bibinfo {author}
  {\bibfnamefont {M.~W.}\ \bibnamefont {Johnson}}, \bibinfo {author}
  {\bibfnamefont {P.}~\bibnamefont {Bunyk}}, \bibinfo {author} {\bibfnamefont
  {E.}~\bibnamefont {Ladizinsky}}, \bibinfo {author} {\bibfnamefont
  {N.}~\bibnamefont {Ladizinsky}}, \bibinfo {author} {\bibfnamefont
  {T.}~\bibnamefont {Oh}}, \ and\ \bibinfo {author} {\bibfnamefont
  {S.}~\bibnamefont {Han}},\ }\href@noop {} {\bibfield  {journal} {\bibinfo
  {journal} {Phys. Rev. B}\ }\textbf {\bibinfo {volume} {80}},\ \bibinfo
  {pages} {052506} (\bibinfo {year} {2009})}\BibitemShut {NoStop}%
\bibitem [{\citenamefont {Tsomokos}\ \emph {et~al.}(2010)\citenamefont
  {Tsomokos}, \citenamefont {Ashhab},\ and\ \citenamefont
  {Nori}}]{tsomokos2010usingsuperconducting}%
  \BibitemOpen
  \bibfield  {author} {\bibinfo {author} {\bibfnamefont {D.~I.}\ \bibnamefont
  {Tsomokos}}, \bibinfo {author} {\bibfnamefont {S.}~\bibnamefont {Ashhab}}, \
  and\ \bibinfo {author} {\bibfnamefont {F.}~\bibnamefont {Nori}},\ }\href@noop
  {} {\bibfield  {journal} {\bibinfo  {journal} {Phys. Rev. A}\ }\textbf
  {\bibinfo {volume} {82}},\ \bibinfo {pages} {052311} (\bibinfo {year}
  {2010})}\BibitemShut {NoStop}%
\bibitem [{\citenamefont {Harris}\ \emph {et~al.}(2010)\citenamefont {Harris},
  \citenamefont {Johnson}, \citenamefont {Lanting}, \citenamefont {Berkley},
  \citenamefont {Johansson}, \citenamefont {Bunyk}, \citenamefont {Tolkacheva},
  \citenamefont {Ladizinsky}, \citenamefont {Ladizinsky}, \citenamefont {Oh}
  \emph {et~al.}}]{harris2010experimental}%
  \BibitemOpen
  \bibfield  {author} {\bibinfo {author} {\bibfnamefont {R.}~\bibnamefont
  {Harris}}, \bibinfo {author} {\bibfnamefont {M.~W.}\ \bibnamefont {Johnson}},
  \bibinfo {author} {\bibfnamefont {T.}~\bibnamefont {Lanting}}, \bibinfo
  {author} {\bibfnamefont {A.~J.}\ \bibnamefont {Berkley}}, \bibinfo {author}
  {\bibfnamefont {J.}~\bibnamefont {Johansson}}, \bibinfo {author}
  {\bibfnamefont {P.}~\bibnamefont {Bunyk}}, \bibinfo {author} {\bibfnamefont
  {E.}~\bibnamefont {Tolkacheva}}, \bibinfo {author} {\bibfnamefont
  {E.}~\bibnamefont {Ladizinsky}}, \bibinfo {author} {\bibfnamefont
  {N.}~\bibnamefont {Ladizinsky}}, \bibinfo {author} {\bibfnamefont
  {T.}~\bibnamefont {Oh}},  \emph {et~al.},\ }\href@noop {} {\bibfield
  {journal} {\bibinfo  {journal} {Phys. Rev. B}\ }\textbf {\bibinfo {volume}
  {82}},\ \bibinfo {pages} {024511} (\bibinfo {year} {2010})}\BibitemShut
  {NoStop}%
\bibitem [{\citenamefont {Johnson}\ \emph {et~al.}(2011)\citenamefont
  {Johnson}, \citenamefont {Amin}, \citenamefont {Gildert}, \citenamefont
  {Lanting}, \citenamefont {Hamze}, \citenamefont {Dickson}, \citenamefont
  {Harris}, \citenamefont {Berkley}, \citenamefont {Johansson}, \citenamefont
  {Bunyk}, \citenamefont {Chapple}, \citenamefont {Enderud}, \citenamefont
  {Hilton}, \citenamefont {Karimi}, \citenamefont {Ladizinsky}, \citenamefont
  {Ladizinsky}, \citenamefont {Oh}, \citenamefont {Perminov}, \citenamefont
  {Rich}, \citenamefont {Thom}, \citenamefont {Tolkacheva}, \citenamefont
  {Truncik}, \citenamefont {Uchaikin}, \citenamefont {Wang}, \citenamefont
  {Wilson},\ and\ \citenamefont {Rose}}]{johnson2011quantum}%
  \BibitemOpen
  \bibfield  {author} {\bibinfo {author} {\bibfnamefont {M.~W.}\ \bibnamefont
  {Johnson}}, \bibinfo {author} {\bibfnamefont {M.~H.~S.}\ \bibnamefont
  {Amin}}, \bibinfo {author} {\bibfnamefont {S.}~\bibnamefont {Gildert}},
  \bibinfo {author} {\bibfnamefont {T.}~\bibnamefont {Lanting}}, \bibinfo
  {author} {\bibfnamefont {F.}~\bibnamefont {Hamze}}, \bibinfo {author}
  {\bibfnamefont {N.}~\bibnamefont {Dickson}}, \bibinfo {author} {\bibfnamefont
  {R.}~\bibnamefont {Harris}}, \bibinfo {author} {\bibfnamefont {A.~J.}\
  \bibnamefont {Berkley}}, \bibinfo {author} {\bibfnamefont {J.}~\bibnamefont
  {Johansson}}, \bibinfo {author} {\bibfnamefont {P.}~\bibnamefont {Bunyk}},
  \bibinfo {author} {\bibfnamefont {E.~M.}\ \bibnamefont {Chapple}}, \bibinfo
  {author} {\bibfnamefont {C.}~\bibnamefont {Enderud}}, \bibinfo {author}
  {\bibfnamefont {J.~P.}\ \bibnamefont {Hilton}}, \bibinfo {author}
  {\bibfnamefont {K.}~\bibnamefont {Karimi}}, \bibinfo {author} {\bibfnamefont
  {E.}~\bibnamefont {Ladizinsky}}, \bibinfo {author} {\bibfnamefont
  {N.}~\bibnamefont {Ladizinsky}}, \bibinfo {author} {\bibfnamefont
  {T.}~\bibnamefont {Oh}}, \bibinfo {author} {\bibfnamefont {I.}~\bibnamefont
  {Perminov}}, \bibinfo {author} {\bibfnamefont {C.}~\bibnamefont {Rich}},
  \bibinfo {author} {\bibfnamefont {M.~C.}\ \bibnamefont {Thom}}, \bibinfo
  {author} {\bibfnamefont {E.}~\bibnamefont {Tolkacheva}}, \bibinfo {author}
  {\bibfnamefont {C.~J.~S.}\ \bibnamefont {Truncik}}, \bibinfo {author}
  {\bibfnamefont {S.}~\bibnamefont {Uchaikin}}, \bibinfo {author}
  {\bibfnamefont {J.}~\bibnamefont {Wang}}, \bibinfo {author} {\bibfnamefont
  {B.}~\bibnamefont {Wilson}}, \ and\ \bibinfo {author} {\bibfnamefont
  {G.}~\bibnamefont {Rose}},\ }\href@noop {} {\bibfield  {journal} {\bibinfo
  {journal} {Nature (London)}\ }\textbf {\bibinfo {volume} {473}},\ \bibinfo
  {pages} {194} (\bibinfo {year} {2011})}\BibitemShut {NoStop}%
\bibitem [{\citenamefont {Grajcar}\ \emph {et~al.}(2005)\citenamefont
  {Grajcar}, \citenamefont {Izmalkov}, \citenamefont {van~der Ploeg},
  \citenamefont {Linzen}, \citenamefont {Ilichev}, \citenamefont {Wagner},
  \citenamefont {H\"{u}bner}, \citenamefont {Meyer}, \citenamefont {Maassen
  van~den Brink}, \citenamefont {Uchaikin},\ and\ \citenamefont
  {Zagoskin}}]{grajcar2005directjosephson}%
  \BibitemOpen
  \bibfield  {author} {\bibinfo {author} {\bibfnamefont {M.}~\bibnamefont
  {Grajcar}}, \bibinfo {author} {\bibfnamefont {A.}~\bibnamefont {Izmalkov}},
  \bibinfo {author} {\bibfnamefont {S.~H.~W.}\ \bibnamefont {van~der Ploeg}},
  \bibinfo {author} {\bibfnamefont {S.}~\bibnamefont {Linzen}}, \bibinfo
  {author} {\bibfnamefont {E.}~\bibnamefont {Ilichev}}, \bibinfo {author}
  {\bibfnamefont {T.}~\bibnamefont {Wagner}}, \bibinfo {author} {\bibfnamefont
  {U.}~\bibnamefont {H\"{u}bner}}, \bibinfo {author} {\bibfnamefont {H.~G.}\
  \bibnamefont {Meyer}}, \bibinfo {author} {\bibfnamefont {A.}~\bibnamefont
  {Maassen van~den Brink}}, \bibinfo {author} {\bibfnamefont {S.}~\bibnamefont
  {Uchaikin}}, \ and\ \bibinfo {author} {\bibfnamefont {A.~M.}\ \bibnamefont
  {Zagoskin}},\ }\href@noop {} {\bibfield  {journal} {\bibinfo  {journal}
  {Phys. Rev. B}\ }\textbf {\bibinfo {volume} {72}},\ \bibinfo {pages} {020503}
  (\bibinfo {year} {2005})}\BibitemShut {NoStop}%
\bibitem [{\citenamefont {Leib}\ and\ \citenamefont
  {Hartmann}(2010)}]{leib2010bosetextendashhubbard}%
  \BibitemOpen
  \bibfield  {author} {\bibinfo {author} {\bibfnamefont {M.}~\bibnamefont
  {Leib}}\ and\ \bibinfo {author} {\bibfnamefont {M.~J.}\ \bibnamefont
  {Hartmann}},\ }\href@noop {} {\bibfield  {journal} {\bibinfo  {journal} {New
  J. Phys.}\ }\textbf {\bibinfo {volume} {12}},\ \bibinfo {pages} {093031}
  (\bibinfo {year} {2010})}\BibitemShut {NoStop}%
\bibitem [{\citenamefont {Makin}\ \emph {et~al.}(2009)\citenamefont {Makin},
  \citenamefont {Cole}, \citenamefont {Hill}, \citenamefont {Greentree},\ and\
  \citenamefont {Hollenberg}}]{makin2009timeevolution}%
  \BibitemOpen
  \bibfield  {author} {\bibinfo {author} {\bibfnamefont {M.~I.}\ \bibnamefont
  {Makin}}, \bibinfo {author} {\bibfnamefont {J.~H.}\ \bibnamefont {Cole}},
  \bibinfo {author} {\bibfnamefont {C.~D.}\ \bibnamefont {Hill}}, \bibinfo
  {author} {\bibfnamefont {A.~D.}\ \bibnamefont {Greentree}}, \ and\ \bibinfo
  {author} {\bibfnamefont {L.~C.~L.}\ \bibnamefont {Hollenberg}},\ }\href@noop
  {} {\bibfield  {journal} {\bibinfo  {journal} {Phys. Rev. A}\ }\textbf
  {\bibinfo {volume} {80}},\ \bibinfo {pages} {043842} (\bibinfo {year}
  {2009})}\BibitemShut {NoStop}%
\bibitem [{\citenamefont {Koch}\ and\ \citenamefont
  {LeHur}(2009)}]{koch2009superfluidmottinsulator}%
  \BibitemOpen
  \bibfield  {author} {\bibinfo {author} {\bibfnamefont {J.}~\bibnamefont
  {Koch}}\ and\ \bibinfo {author} {\bibfnamefont {K.}~\bibnamefont {LeHur}},\
  }\href@noop {} {\bibfield  {journal} {\bibinfo  {journal} {Phys. Rev. A}\
  }\textbf {\bibinfo {volume} {80}},\ \bibinfo {pages} {023811} (\bibinfo
  {year} {2009})}\BibitemShut {NoStop}%
\bibitem [{\citenamefont {Schmidt}\ \emph {et~al.}(2010)\citenamefont
  {Schmidt}, \citenamefont {Gerace}, \citenamefont {Houck}, \citenamefont
  {Blatter},\ and\ \citenamefont {T\"{u}reci}}]{schmidt2010nonequilibrium}%
  \BibitemOpen
  \bibfield  {author} {\bibinfo {author} {\bibfnamefont {S.}~\bibnamefont
  {Schmidt}}, \bibinfo {author} {\bibfnamefont {D.}~\bibnamefont {Gerace}},
  \bibinfo {author} {\bibfnamefont {A.~A.}\ \bibnamefont {Houck}}, \bibinfo
  {author} {\bibfnamefont {G.}~\bibnamefont {Blatter}}, \ and\ \bibinfo
  {author} {\bibfnamefont {H.~E.}\ \bibnamefont {T\"{u}reci}},\ }\href@noop {}
  {\bibfield  {journal} {\bibinfo  {journal} {Phys. Rev. B}\ }\textbf {\bibinfo
  {volume} {82}},\ \bibinfo {pages} {100507} (\bibinfo {year}
  {2010})}\BibitemShut {NoStop}%
\bibitem [{\citenamefont {Sachdev}(2011)}]{sachdev2011quantum}%
  \BibitemOpen
  \bibfield  {author} {\bibinfo {author} {\bibfnamefont {S.}~\bibnamefont
  {Sachdev}},\ }\href@noop {} {\emph {\bibinfo {title} {Quantum Phase
  Transitions}}}\ (\bibinfo  {publisher} {Cambridge University Press},\
  \bibinfo {year} {2011})\BibitemShut {NoStop}%
\bibitem [{\citenamefont {Scala}\ \emph {et~al.}(2007)\citenamefont {Scala},
  \citenamefont {Militello}, \citenamefont {Messina}, \citenamefont
  {Maniscalco}, \citenamefont {Piilo},\ and\ \citenamefont
  {Suominen}}]{scala2007cavitylosses}%
  \BibitemOpen
  \bibfield  {author} {\bibinfo {author} {\bibfnamefont {M.}~\bibnamefont
  {Scala}}, \bibinfo {author} {\bibfnamefont {B.}~\bibnamefont {Militello}},
  \bibinfo {author} {\bibfnamefont {A.}~\bibnamefont {Messina}}, \bibinfo
  {author} {\bibfnamefont {S.}~\bibnamefont {Maniscalco}}, \bibinfo {author}
  {\bibfnamefont {J.}~\bibnamefont {Piilo}}, \ and\ \bibinfo {author}
  {\bibfnamefont {K.}~\bibnamefont {Suominen}},\ }\href@noop {} {\bibfield
  {journal} {\bibinfo  {journal} {J. Phys. A: Math. Theor.}\ }\textbf {\bibinfo
  {volume} {40}},\ \bibinfo {pages} {14527} (\bibinfo {year}
  {2007})}\BibitemShut {NoStop}%
\bibitem [{\citenamefont {Reuther}\ \emph {et~al.}(2010)\citenamefont
  {Reuther}, \citenamefont {Zueco}, \citenamefont {Deppe}, \citenamefont
  {Hoffmann}, \citenamefont {Menzel}, \citenamefont {Weissl}, \citenamefont
  {Mariantoni}, \citenamefont {Kohler}, \citenamefont {Marx}, \citenamefont
  {Solano} \emph {et~al.}}]{reuther2010tworesonator}%
  \BibitemOpen
  \bibfield  {author} {\bibinfo {author} {\bibfnamefont {G.~M.}\ \bibnamefont
  {Reuther}}, \bibinfo {author} {\bibfnamefont {D.}~\bibnamefont {Zueco}},
  \bibinfo {author} {\bibfnamefont {F.}~\bibnamefont {Deppe}}, \bibinfo
  {author} {\bibfnamefont {E.}~\bibnamefont {Hoffmann}}, \bibinfo {author}
  {\bibfnamefont {E.~P.}\ \bibnamefont {Menzel}}, \bibinfo {author}
  {\bibfnamefont {T.}~\bibnamefont {Weissl}}, \bibinfo {author} {\bibfnamefont
  {M.}~\bibnamefont {Mariantoni}}, \bibinfo {author} {\bibfnamefont
  {S.}~\bibnamefont {Kohler}}, \bibinfo {author} {\bibfnamefont
  {A.}~\bibnamefont {Marx}}, \bibinfo {author} {\bibfnamefont {E.}~\bibnamefont
  {Solano}},  \emph {et~al.},\ }\href@noop {} {\bibfield  {journal} {\bibinfo
  {journal} {Phys. Rev. B}\ }\textbf {\bibinfo {volume} {81}},\ \bibinfo
  {pages} {144510} (\bibinfo {year} {2010})}\BibitemShut {NoStop}%
\end{thebibliography}

\end{document}